\documentclass[showpacs,preprintnumbers,amsmath,amssymb,nofootinbib,floatfix]{revtex4}

\usepackage[dvips]{graphicx}
\usepackage{dcolumn}
\usepackage{bm}

\def\mapgeq{\mathbin{\lower.3ex\hbox{$\buildrel>\over{\smash{\scriptstyle\sim}\vphantom{_x}}$}}}
\def\mapleq{\mathbin{\lower.3ex\hbox{$\buildrel<\over{\smash{\scriptstyle\sim}\vphantom{_x}}$}}}
\def\mapgeqeq{\mathbi{\lower.3ex\hbox{$\buildrel>\over{\smash{\scriptstyle\approx}\vphantom{_2}}$}}}
\def\mapleqeq{\mathbin{\lower.3ex\hbox{$\buildrel<\over{\smash{\scriptstyle\approx}\vphantom{_2}}$}}}
\def\Journal#1#2#3#4{{#1} {\bf #2} (#4) #3}

\def\NPB{Nucl. Phys. B}

\def\NPSUPPL{Nucl. Phys. Proc. Suppl.}
\def\PLB{{Phys. Lett.} B}

\def\PLBOLD{Phys. Lett.}
\def\PRL{Phys. Rev. Lett.}

\def\PRD{Phys. Rev. D}

\def\PTP{Prog. Theor. Phys.}
\def\JHEP{JHEP}

\def\JETPUSSR{Sov. Phys. JETP}
\def\JETPUSSRLETT{Sov. Phys. JETP Letters}
\def\ZETP{Zh. Eksp. Teor. Fiz.}
\def\PismaZETP{Pis'ma Zh. Eksp. Teor. Fiz.}

\def\IJMPE{Int. J. Mod. Phys. E}
\def\JPG{J. Phys. G}

\def\ACTAB{Acta. Phys. Pol. B}
\def\Erratum{Erratum-ibid}

\begin{document}

\preprint{TOKAI-HEP/TH-0701}

\title{Leptonic CP Violation Induced by Approximately $\mu$-$\tau$ Symmetric Seesaw Mechanism}

\author{Teppei Baba}
\email{7atrd014@keyaki.cc.u-tokai.ac.jp}
\author{Masaki Yasu\`{e}}%
\email{yasue@keyaki.cc.u-tokai.ac.jp}
\affiliation{\vspace{3mm}%
\sl Department of Physics, Tokai University,\\
1117 Kitakaname, Hiratsuka, Kanagawa 259-1292, Japan\\
}

\date{October, 2007}

\begin{abstract}
Assuming a minimal seesaw model with two heavy neutrinos ($N$), we examine effects of leptonic CP violation 
induced by approximate $\mu$-$\tau$ symmetric interactions.  As long as $N$ is 
subject to the $\mu$-$\tau$ symmetry, 
we can choose CP phases of Dirac mass terms without loss of generality in such 
a way that these phases arise from $\mu$-$\tau$ symmetry breaking interactions. In the case that 
no phase is present in heavy neutrino mass terms, leptonic CP phases are controlled by 
two phases $\alpha$ and $\beta$. 
The similar consideration is extended to $N$ blind to the $\mu$-$\tau$ symmetry.
 It is argued that $N$ subject (blind) to the $\mu$-$\tau$ symmetry 
necessarily describes the normal (inverted) mass hierarchy. 
We restrict ourselves to $\mu$-$\tau$ symmetric textures giving the tri-bimaximal mixing
and calculate flavor neutrino masses to estimate CP-violating Dirac and Majorana phases as well as 
neutrino mixing angles as functions of $\alpha$ and $\beta$. 
Since $\alpha$ and $\beta$ are generated by $\mu$-$\tau$ 
symmetry breaking interactions, CP-violating Majorana phase tends to be 
suppressed and is found to be at most ${\mathcal O}(0.1)$ radian.  On the other hand, CP-violating Dirac phase 
tends to show a proportionality to $\alpha$ or to $\beta$.
\end{abstract}

\maketitle
\section{\label{sec:1}Introduction}
Recent extensive analysis on neutrino oscillations \cite{NeutrinoData} has indicated almost maximal atmospheric 
neutrino mixing and large solar neutrino mixing as well as suppressed reactor neutrino mixing angle. 
These observed properties can well be understood by assuming a $\mu$-$\tau$ 
symmetry in neutrino interactions \cite{mu-tau}.  Another interesting property of neutrinos, which has 
not yet been observed, is related to leptonic CP violation. The leptonic CP violation of 
the Dirac type is known to be absent in the $\mu$-$\tau$ symmetric limit \cite{mu-tauCP}.  Therefore, 
to discuss physics of leptonic CP violation needs the $\mu$-$\tau$ symmetry breaking in neutrino interactions. 

Leptonic CP violation can be parameterized by one Dirac CP-violating phase ($\delta_{CP}$) and 
three Majorana phases ($\phi_{1,2,3}$) \cite{CPphases} in the Pontecorvo-Maki-Nakagawa-Sakata 
(PMNS) mixing matrix $U_{PMNS}=U^0_\nu K^0$ \cite{PMNS} with
\begin{eqnarray}
U^0_\nu&=&\left( \begin{array}{ccc}
  c_{12}c_{13} &  s_{12}c_{13}&  s_{13}e^{-i\delta_{CP}}\\
  -c_{23}s_{12}-s_{23}c_{12}s_{13}e^{i\delta_{CP}}
                                 &  c_{23}c_{12}-s_{23}s_{12}s_{13}e^{i\delta_{CP}}
                                 &  s_{23}c_{13}\\
  s_{23}s_{12}-c_{23}c_{12}s_{13}e^{i\delta_{CP}}
                                 &  -s_{23}c_{12}-c_{23}s_{12}s_{13}e^{i\delta_{CP}}
                                 & c_{23}c_{13},\\
\end{array} \right),
\nonumber \\
K^0 &=& {\rm diag}(e^{i\phi_1}, e^{i\phi_2}, e^{i\phi_3}),
\label{Eq:Uu}
\end{eqnarray}
where $c_{ij}=\cos\theta_{ij}$ and $s_{ij}=\sin\theta_{ij}$ ($i,j$=1,2,3), as adopted by the 
Particle Data Group \cite{PDG}. 
The Majorana CP-violating phases are determined by two combinations of $\phi_{1,2,3}$ such 
as $\phi_i-\phi_1$ ($i$=1,2,3).  Since there are arbitrary phases of the flavor neutrinos, 
the phases of $U_{PMNS}$ vary with these phases.  The most general 
form of $U_{PMNS}$ is given by 
$U_\nu$ and $K$ \cite{BabaYasue} in place of $U^0_\nu$ and $K^0$:
\begin{eqnarray}
U_\nu&=&\left( {\begin{array}{*{20}c}
   1 & 0 & 0  \\
   0 & {e^{i\gamma } } & 0  \\
   0 & 0 & {e^{ - i\gamma } }  \\
\end{array}} \right)\left( {\begin{array}{*{20}c}
   {c_{12} c_{13} } & {s_{12} c_{13} e^{i\rho } } & {s_{13} e^{ - i\delta } }  \\
   { - c_{23} s_{12} e^{ - i\rho }  - s_{23} c_{12} s_{13} e^{i\delta } } & {c_{23} c_{12}  - s_{23} s_{12} s_{13} e^{i\left( {\delta  + \rho } \right)} } & {s_{23} c_{13} }  \\
   {s_{23} s_{12} e^{ - i\rho }  - c_{23} c_{12} s_{13} e^{i\delta } } & { - s_{23} c_{12}  - c_{23} s_{12} s_{13} e^{i\left( {\delta  + \rho } \right)} } & {c_{23} c_{13} }  \\
\end{array}} \right),
\nonumber \\
K &=& {\rm diag}(e^{i\phi^\prime_1}, e^{i\phi_2}, e^{i\phi_3}),
\label{Eq:Uu_general}
\end{eqnarray}
where $\delta_{CP}=\delta+\rho$ and $\phi_1 = \phi^\prime_1-\rho$, 
which will be used in this article.\footnote{The phases $\rho$ and $\gamma$ 
are redundant and can be removed by the redefinition of flavor neutrino masses.  
The resultant $U_{PMNS}$ involves 
$\delta+\rho$ and $ \phi^\prime_1-\rho$, respectively, as Dirac and Majorana CP-violating phases.  
Although Eq.(\ref{Eq:Uu_general}) contains 6 CP phases, strictly speaking, there 
are 7 CP phases in total \cite{MostGeneral}.  In fact, there is an additional phase for 
the $2-3$ rotation ($\tau$) contributing to $\delta_{CP}$ as $\delta_{CP} = \delta+\rho+\tau$, 
which, however, can be removed by introducing a new definition: $\rho^\prime=\rho+\tau/2$,
$\gamma^\prime=\gamma+\tau/2$ and
 $\delta^\prime=\delta+\tau/2$.  As a reault, we end up with the same definition of 
$\delta_{CP}$: $\delta_{CP}=\delta^\prime + \rho^\prime$. See Ref.\cite{BabaYasue} 
for more details. 
Therefore, the parameterization with 
$\delta^\prime$, $\rho^\prime$, and $\gamma^\prime$ gives 
Eq.(\ref{Eq:Uu_general}) as a general form of $U_{PMNS}$.}
Another aspect of the role of leptonic CP phases 
may lie in creation of the baryon asymmetry of the Universe \cite{CP-Baryon} 
when the seesaw mechanism active at higher energies \cite{leptogenesis} is responsible for 
generating neutrino masses \cite{MostGeneral,SeeSaw}.  
More precisely, the minimal seesaw mechanism based 
on two heavy neutrinos ($N$) provides the direct linkage 
between the high energy phases in the seesaw mechanism and the low energy phases defined by $U_{PMNS}$ \cite{TwoHeavy}.
Therefore, we can predict the size of the low energy phases that 
yields the observed size of the baryon asymmetry of the Universe.

In the present article, we discuss 
$\mu$-$\tau$ symmetry breaking effects in the minimal seesaw mechanism 
on flavor neutrino masses to study low energy CP violation \cite{mu-tauSeeSaw}. The correlation between 
the low energy CP violation and its effect in leptogenesis will be discussed in a subsequent article. 
In Sec.\ref{sec:2}, our minimal seesaw model 
is described.  In Sec.\ref{sec:3}, we calculate flavor neutrino masses in models based on $N$ subject to 
the $\mu$-$\tau$ symmetry where the normal mass hierarchy is realized
 and on $N$ blind to the $\mu$-$\tau$ symmetry where the inverted mass hierarchy is realized.
How these mass hierarchies arise is discussed in the Appendix \ref{sec:Appendix1}.
We restrict ourselves 
to mass terms of $N$ without CP-violating phases.  As a result, 
leptonic CP violation arises in general 
from the $\mu$-$\tau$ symmetry breaking terms for the normal mass hierarchy.  
To compare our predictions 
with those for the inverted mass hierarchy, the same phase structure is assumed 
for the inverted mass hierarchy.
To simplify our discussions, the $\mu$-$\tau$ symmetric flavor neutrino 
mass texture is taken to describe the tri-bimaximal mixing \cite{TriBiMaximal}, 
which predicts the consistent value of $\sin\theta_{12}$ with the observed data.
We choose three textures to describe the normal and inverted mass hierarchies.  
The calculation of the flavor neutrino masses supplied by the seesaw mechanism 
is performed in the Appendix \ref{sec:Appendix1}.
In Sec.\ref{sec:4}, CP-violating phases as well as neutrino masses and mixing angles are 
calculated and results are shown in figures.  A set of formula used in Sec.\ref{sec:4} is 
summarized in the Appendix \ref{sec:Appendix2}.  Final section is devoted to summary and discussions.

\section{\label{sec:2}Model}
Let us begin with defining superpotential for leptons ($W$) 
in the minimal seesaw model with three flavors of $L$ ($\ell$) as 
$SU(2)_L$-doublets (singlets) and two flavors of $N$ to be denoted by ($N_\mu$, $N_\tau$) as $SU(2)_L$-singlets 
as well as two Higgses $H_{u,d}$:
\begin{eqnarray}
&& 
W = e^{CT} {\bf Y}_\ell  LH_d  + N^{CT} {\bf Y}_\nu  LH_u  + \frac{1}{2}N^{CT} {\bf M}_R N^C 
\label{Eq:SuperPotential}
\end{eqnarray}
where ${\bf Y}_\ell$ and ${\bf Y}_\nu$ are Yukawa couplings and ${\bf M}_R$ is a Majorana mass matrix of $N$.  
We can always choose the base, where ${\bf Y}_\ell$ is diagonal, which defines the charged leptons 
$e$, $\mu$ and $\tau$. 
The coupling ${\bf Y}_\nu$ and the mass matrix ${\bf M}_R$ are parameterized as follows:
\begin{eqnarray}
&& 
{\bf Y}_\nu   = \left( {\begin{array}{*{20}c}
   {h_{\mu e} } \hfill & {h_{\mu \mu } } \hfill & {h_{\mu \tau } } \hfill  \\
   {h_{\tau e} } \hfill & {h_{\tau \mu } } \hfill & {h_{\tau \tau } } \hfill  \\
\end{array}} \right),
\quad
{\bf M}_R  = \left( {\begin{array}{*{20}c}
   {{\bf M}_{R\mu \mu } } & {{\bf M}_{R\mu \tau } }  \\
   {{\bf M}_{R\mu \tau } } & {{\bf M}_{R\tau \tau } }  \\
\end{array}} \right).
\label{Eq:MassMatrix}
\end{eqnarray}
Our $\mu$-$\tau$ symmetry is defined by the invariance of $W$ under the interchange 
$\nu_\mu\leftrightarrow -\sigma\nu_\tau$, 
where $\sigma = \pm 1$ will take care of the sign of $\sin\theta_{23}$.\footnote{This interchange should be 
replaced by $L_\mu\leftrightarrow -\sigma L_\tau$ 
and is consistently described if two extra Higgses $H^\prime_{u,d}$ are introduced. 
Under the interchange, we require that 
$H^\prime_{u,d}\rightarrow -H^\prime_{u,d}$ while 
$H_{u,d}\rightarrow H_{u,d}$, where $\langle 0 \vert H^\prime_{u,d} \vert 0\rangle$ 
yields $\mu$-$\tau$ symmetry breaking terms.  The interplay of these 
Higgses supplies a $\mu$-$\tau$ symmetric $W$ and also accounts for the appearance of the 
badly broken $\mu$-$\tau$ symmetry for the 
charged leptons and of the approximate $\mu$-$\tau$ symmetry for neutrinos \cite{Charged mu-tau}.  For the purpose 
of the present article, 
it is sufficient to use Eq.(\ref{Eq:SuperPotential}).} 
This form of the interchange is based on the 
choice of $\sin\theta_{23}=\sigma/\sqrt{2}$ defined in $U_{PMNS}$ of Eq.(\ref{Eq:Uu}) in the 
$\mu$-$\tau$ symmetric limit. It is readily seen that the corresponding mass matrix is
$M^{(+)}$ of Eq.(\ref{Eq:Mnu-mutau-separation-2}), which has $(0, \sigma/\sqrt{2}, 1/\sqrt{2})^T$ 
as an eigenvector that in turn gives $\sin\theta_{23}=\sigma/\sqrt{2}$ from the third column of 
$U_{PMNS}$ as long as this eigenvector is assigned to the third neutrino \cite{13-12}.  As a 
result, $M^{(+)}$ is invariant under $\nu_\mu\leftrightarrow -\sigma\nu_\tau$.  It is 
convenient to introduce 
\begin{eqnarray}
&& 
\nu _ \pm   = \frac{{\nu _\mu   \pm \left( { - \sigma \nu _\tau  } \right)}}{{\sqrt 2 }},
\quad
N _ \pm^C   = \frac{{N _\mu^C   \pm \left( { - \sigma N _\tau^C  } \right)}}{{\sqrt 2 }},
\label{Eq:NuPlusMinus}
\end{eqnarray}
to discuss property of neutrinos with respect to the $\mu$-$\tau$ symmetry.  

In addition to the assignment of $N$ to be $N=(N_\mu, N_\tau)$, there are other cases, 
which contain $N_e$ in models such as those based on $(N_e, N_{\mu,\tau})$.  For $N$ subject to the 
$\mu$-$\tau$ symmetry, $N$ can be either $(N_+, N_-)$ or $(N_e, N_-)$ while, for $N$ blind to the 
$\mu$-$\tau$ symmetry, $N_\pm$ can be any two heavy neutrinos of $N_e$, $N_\mu$ and $N_\tau$.
To treat these cases, we use $N=(N_+, N_-)$ instead of $N=(N_\mu, N_\tau)$.   
The coupling ${\bf Y}_\nu$ and the mass matrix ${\bf M}_R$  are parameterized as follows:
\begin{eqnarray}
&& 
{\bf Y}_\nu   = \left( {\begin{array}{*{20}c}
   {h_{ + e} } \hfill & {h_{ + \mu } } \hfill & {h_{ + \tau } } \hfill  \\
   {h_{ - e} } \hfill & {h_{ - \mu } } \hfill & {h_{ - \tau } } \hfill  \\
\end{array}} \right),
\quad
{\bf M}_R  = \left( {\begin{array}{*{20}c}
   {{\bf M}_{R++ } } & {{\bf M}_{R+- } }  \\
   {{\bf M}_{R+- } } & {{\bf M}_{R-- } }  \\
\end{array}} \right).
\label{Eq:MassMatrix-e}
\end{eqnarray}
Another case with $N$ subject to the $\mu$-$\tau$ symmetry can be discussed by 
replacing $(N_+, N_-)$ with $(N_e, N_-)$.  We classify all cases
 by specifying couplings of the Yukawa interactions for neutrinos
 denoted by $f$'s, where neutrinos are expressed in terms of 
 $N_\pm$ and $\nu_{e, \pm}$. The corresponding lagrangian ${\mathcal L}_\nu$ is 
 described by
\begin{eqnarray}
-{\mathcal L}_\nu
&=& 
\left( {f_{ e e}{\overline {N_ e}} + f_{ + e}{\overline {N_ +}}  + f_{ - e} {\overline {N_ -}} } \right)\nu _e  H_{u1}
+ \left( {f_{ e  + }\overline {N_ e}  + f_{ +  + }\overline {N_ +}  + f_{ -  + } {\overline {N_ -}} } \right)\nu _ +   H_{u1} 
\nonumber\\
&+& 
\left( {f_{ e  - } \overline {N_ e} +f_{ +  - } {\overline {N_ +}}  +f_{ -  - } \overline {N_ -}  } \right)\nu _ -   H_{u1} .
\label{Eq:FlavorCoulingPlusMinus}
\end{eqnarray}
where $H_{u1}$ is defined in $H_u = (H_{u1}, H_{u2})^T$.  In each case, if $N$ is subject to the $\mu$-$\tau$ symmetry,
$N_+$ can be $(N_\mu+(-\sigma)N_\tau)/\sqrt{2}$ or $N_e$ and 
$N_-$ is just  $(N_\mu-(-\sigma)N_\tau)/\sqrt{2}$.  
On the other hand, if $N$ is blind to the $\mu$-$\tau$ symmetry, $N_\pm$ can be any combinations of $N_{e,\mu,\tau}$.
 The difference of these assignments is absorbed into the definition of $f$'s, 
 which are given by different Yukawa couplings in the starting superpotential expressed 
 in terms of $N_{e,\mu,\tau}$ and $\nu_{e,\mu,\tau}$.\footnote{In a seesaw model with $N_{e,\mu,\tau}$, 
 two light heavy neutrinos that effectively describes the minimal seesaw model 
 discussed here are dynamically determined by mass terms of $N_{e,\mu,\tau}$ and can be 
 any combination of $N_{e,\mu,\tau}$.  However, if $N$ in the minimal seesaw model 
 is subject to the $\mu$-$\tau$ symmetry, $N$ should include $N_-$ as a light heavy neutrino.} 
 It is sufficient to use the notation of $N_\pm$ for the later discussions.

For a complex ${\bf M}_R$, phases of $\theta_{Rij}$ defined by 
${\bf M}_{Rij}= \exp(i\theta_{Rij}){\bf M}_{ij}$ ($i.j=+.-$)
 can be transferred into the phase of 
${\bf M}_{R+-}$, 
where ${\bf M}_{ij}$ ($i,j=+,-$) are taken to be real for $-\pi/2\leq\theta_{Rij}\leq\pi/2$. 
We here use ($+$, $-$) as the suffix of {\bf M}, which 
should be replaced by other combinations such as ($e$, $-$), appropriately. 
This phase becomes $\theta_{R +-}-(\theta_{R ++}+\theta_{R --})/2(\equiv \Theta_{+-})$ 
and ${\bf M}_R$ is given by
\begin{eqnarray}
&& 
{\bf M}_R  = \left( {\begin{array}{*{20}c}
   {{\bf M}_{+ + } } & e^{i\Theta_{+-}}{ {\bf M}_{+ - } }  \\
   e^{i\Theta_{+-}}{{\bf M}_{+ - } } & {{\bf M}_{- - } }  \\
\end{array}} \right).
\label{Eq:HeavyMassMatrix}
\end{eqnarray}
Without the loss of generality, we choose that ${\bf M}_{-- } > {\bf M}_{+ + }$.
The unitary matrix $U$ that diagonalizes ${\bf M}_R$ to give 
\begin{eqnarray}
&& 
{\bf M}_{{\rm{diag}}}  = U^ *  {\bf M}_R U^\dag
=
\left( {\begin{array}{*{20}c}
   {M_ 1e^{2i\varphi _ 1  }   } & 0  \\
   0 & {M_ 2e^{2i\varphi _ 2  }   }  \\
\end{array}} \right)
\label{Eq:HeavyUnitaryMatrix}
\end{eqnarray}
is 
\begin{eqnarray}
&& 
U  = \left( {\begin{array}{*{20}c}
   \cos\theta & { - e^{i\omega } \sin\theta}  \\
   {e^{ - i\omega } \sin\theta} & \cos\theta  \\
\end{array}} \right),
\label{Eq:HeavyUnitaryMatrix2}
\end{eqnarray}
where  
\begin{eqnarray}
&& 
\omega = \arg\left(e^{i\Theta _{+ -} } {\bf M}_{+ + }  + e^{ - i\Theta _{+ -}}  {\bf M}_{- - }\right),
\label{Eq:HeavyPhaseOmega}\\
&&
 M_ 1  e^{2i\varphi _ 1  }  = \cos^2\theta {\bf M}_{+ + }  + e^{ - 2i\omega } \sin^2\theta {\bf M}_{- - }  - 2\cos\theta\sin\theta e^{i\left( {\Theta _{+ - }  - \omega } \right)} {\bf M}_{+ - }.
\nonumber\\
&&
 M_ 2  e^{2i\varphi _ 2  }  = \cos^2\theta {\bf M}_{- - }  + e^{2i\omega } \sin^2\theta {\bf M}_{+ + }  + 2\cos\theta\sin\theta e^{i\left( {\Theta _{+ - }  + \omega } \right)} {\bf M}_{+ - },
\nonumber\\
\label{Eq:HeavyPhaseMajirana}
\end{eqnarray}
and
\begin{eqnarray}
\tan \theta  = \frac{{2{\bf M}_{ +  - } r}}{{{\bf M}_{ -  - }  - {\bf M}_{ +  + }  + \sqrt {\left( {{\bf M}_{ -  - }  - {\bf M}_{ +  + } } \right)^2  + 4{\bf M}_{ +  - }^2 r^2 } }},
\label{Eq:HeavyMixingAngle}
\end{eqnarray}
The parameter $r$ is given by
\begin{eqnarray}
&& 
r = {\frac{\left|{e^{i\Theta _{+ -} } {\bf M}_{+ + }  + e^{ - i\Theta _{+ -} } {\bf M}_{- - } } \right|}{{{\bf M}_{+ + }  + {\bf M}_{- - } }}}.
\label{Eq:HeavyMixingAngle-r}
\end{eqnarray}
The phase $\omega$ is further expressed as
\begin{eqnarray}
&& 
\tan \omega  = \frac{{\bf M}_{+ + }-{\bf M}_{- - } }
{{\bf M}_{+ + }+{\bf M}_{- - } }\tan \Theta _{+-}.
\label{Eq:HeavyUnitaryMatrixPhase-2}
\end{eqnarray}
In the case of $N_-\rightarrow -N_-$ under the $\mu$-$\tau$ symmetry transformation, we obtain that 
\begin{eqnarray}
&& 
{\bf M}_{+-}=0,
\label{Eq:HeavyUnitaryMatrixSymmetric}
\end{eqnarray}
leading to $\theta = 0$ in the $\mu$-$\tau$ symmetric limit.

The coupling ${\bf Y}_\nu$ for $(N_+,N_-)$ and $(\nu_e,\nu_\mu,\nu_\tau)$ can be parameterized to be:
\begin{eqnarray}
&& 
 {\bf Y}_\nu  = \left( {\begin{array}{*{20}c}
   h_{ + e} \hfill & {h_{ + \mu } } \hfill & { h_{ + \tau } } \hfill  \\
   h_{ - e} \hfill & {h_{ - \mu } } \hfill & { h_{ - \tau } } \hfill  \\
\end{array}} \right),
\label{Eq:Yukawas}
\end{eqnarray}
and formally divided into two parts as 
${\bf Y}_\nu={\bf Y}^{(+)}_\nu+{\bf Y}^{(-)}_\nu$, which is just an identity, where 
 the superscripts $(+)$ and $(-)$ of ${\bf Y}_\nu$ are, respectively, so chosen to stand for the $\mu-\tau$ 
 symmetry preserving and breaking terms.  
 We obtain the following ${\bf Y}_\nu$:
\begin{enumerate}
\item For $N$ subject to the $\mu$-$\tau$ symmetry, 
\begin{eqnarray}
&& 
 {\bf Y}_\nu ^{\left(  +  \right)}  = \left( {\begin{array}{*{20}c}
   {h_{ + e}^{\left(  +  \right)} } \hfill & {h_{ + \mu }^{\left(  +  \right)} } \hfill & { - \sigma h_{ + \mu }^{\left(  +  \right)} } \hfill  \\
   0 \hfill & {h_{ - \mu }^{\left(  +  \right)} } \hfill & {\sigma h_{ - \mu }^{\left(  +  \right)} } \hfill  \\
\end{array}} \right), 
\quad
 {\bf Y}_\nu ^{\left(  -  \right)}  = \left( {\begin{array}{*{20}c}
   0 \hfill & {h_{ + \mu }^{\left(  -  \right)} } \hfill & {\sigma h_{ + \mu }^{\left(  -  \right)} } \hfill  \\
   {h_{ - e}^{\left(  -  \right)} } \hfill & {h_{ - \mu }^{\left(  -  \right)} } \hfill & { - \sigma h_{ - \mu }^{\left(  -  \right)} } \hfill  \\
\end{array}} \right), 
\label{Eq:mu-tauMassMatrix-1}
\end{eqnarray}
where
\begin{eqnarray}
&&
 h_{ + e}^{\left(  +  \right)}  = f_{ + e} , 
 \quad
 h_{ + \mu }^{\left(  +  \right)}  = \frac{{f_{ +  + }}}{{\sqrt 2 }}, 
  \quad
h_{ - \mu }^{\left(  +  \right)}  = \frac{{f_{ -  - } }}{{\sqrt 2 }},
\nonumber\\
&&
 h_{ - e}^{\left(  -  \right)}  = f_{ - e}, 
  \quad
h_{ + \mu }^{\left(  -  \right)}  = \frac{{f_{ +  - } }}{{\sqrt 2 }}, 
 \quad
h_{ - \mu }^{\left(  -  \right)}  = \frac{{f_{ -  + } }}{{\sqrt 2 }},
\label{Eq:mu-tau-h-1}
\end{eqnarray}
from
\begin{eqnarray}
&& 
-{\mathcal L}_\nu
=
\left( {f_{ + e} {\overline {N_ +}}  + f_{ - e} {\overline {N_ -}} } \right)\nu _e  H_{u1}
+ \left( {f_{ +  + }\overline {N_ +}  + f_{ -  + } {\overline {N_ -}} } \right)\nu _ +   H_{u1} 
+ \left( {f_{ +  - } {\overline {N_ +}} + f_{ -  - } \overline {N_ -}   } \right)\nu _ -   H_{u1} ,
\label{Eq:FlavorCoulingPlusMinus-1}
\end{eqnarray}
for $N=(N_+, N_-)$, and where
\begin{eqnarray}
&&
 h_{ + e}^{\left(  +  \right)}  = f_{ e e}, 
  \quad
h_{ + \mu }^{\left(  +  \right)}  = \frac{{f_{ e  + } }}{{\sqrt 2 }}, 
 \quad
h_{ - \mu }^{\left(  +  \right)}  = \frac{{f_{ -  - } }}{{\sqrt 2 }},
\nonumber\\
&&
 h_{ - e}^{\left(  -  \right)}  = f_{ - e} , 
  \quad
h_{ + \mu }^{\left(  -  \right)}  = \frac{{f_{ e  - } }}{{\sqrt 2 }}, 
 \quad
h_{ - \mu }^{\left(  -  \right)}  = \frac{{f_{ -  + } }}{{\sqrt 2 }},
\label{Eq:mu-tau-h-1-2}
\end{eqnarray}
from
\begin{eqnarray}
&& 
-{\mathcal L}_\nu
=
\left( {f_{ee} {\overline {N_ e}}  + f_{ - e} {\overline {N_ -}} } \right)\nu _e  H_{u1}
+ \left( {f_{e + } {\overline {N_ e}}  + f_{ -  + } {\overline {N_ -}} } \right)\nu _ + H_{u1}
+ \left( {f_{ e  - } {\overline {N_ e}} + f_{- - } {\overline {N_ -}}  } \right)\nu _ -  H_{u1},
\label{Eq:FlavorCoulingPlusMinus-1-e}
\end{eqnarray}
for $N=(N_e, N_-)$.  In the $\mu$-$\tau$ symmetric limit, 
it is obvious to see that this case provides $\nu_e\nu_+$, $\nu_+\nu_+$ and $\nu_-\nu_-$ 
as flavor neutrino mass terms and the phase of $U$ is absent because of 
Eq.(\ref{Eq:HeavyUnitaryMatrixSymmetric}).  
Since the quantity $U^ *{\bf Y}^{(+)}_\nu {\bf Y}^{(+)\dagger}_\nu U^T $
related to the leptogenesis turns out 
to be real, the leptogenesis requires $\mu$-$\tau$ symmetry breaking couplings \cite{mu-tauNoCP}.
In the Yukawa interactions, we see that phases of 
($f_{ + e} $, $f_{ +  + }$, $f_{ -  - }$) in Eq.(\ref{Eq:FlavorCoulingPlusMinus-1})
or
($f_{ e e}$, $f_{ e  + }$, $f_{ -  - }$) in Eq.(\ref{Eq:FlavorCoulingPlusMinus-1-e})
are, respectively, absorbed by adjusting phases of $\nu _e$, $\nu _ + $ and $\nu _ -$.  As a result, the phases arises  solely from 
the $\mu$-$\tau$ symmetry breaking couplings.
\item For $N$ blind to the $\mu$-$\tau$ symmetry, 
\begin{eqnarray}
&& 
 {\bf Y}_\nu^{\left(  +  \right)}  = \left( {\begin{array}{*{20}c}
   {h_{ + e}^{\left(  +  \right)} } \hfill & {h_{ + \mu }^{\left(  +  \right)} } \hfill & { - \sigma h_{ + \mu }^{\left(  +  \right)} } \hfill  \\
   {h_{ - e}^{\left(  +  \right)} } \hfill & {h_{ - \mu }^{\left(  +  \right)} } \hfill & { - \sigma h_{ - \mu }^{\left(  +  \right)} } \hfill  \\
\end{array}} \right),
\quad
 {\bf Y}_\nu ^{\left(  -  \right)}  = \left( {\begin{array}{*{20}c}
   0 \hfill & {h_{ + \mu }^{\left(  -  \right)} } \hfill & {\sigma h_{ + \mu }^{\left(  -  \right)} } \hfill  \\
   0 \hfill & {h_{ - \mu }^{\left(  -  \right)} } \hfill & {\sigma h_{ - \mu }^{\left(  -  \right)} } \hfill  \\
\end{array}} \right),
\label{Eq:mu-tauMassMatrix-2}
\end{eqnarray}
where
\begin{eqnarray}
&&
 h_{ + e}^{\left(  +  \right)}  = f_{ + e} , 
  \quad
h_{ - e}^{\left(  +  \right)}  = f_{ - e} , 
 \quad
h_{ + \mu }^{\left(  +  \right)}  = \frac{{f_{ +  + } }}{{\sqrt 2 }}, 
 \quad
h_{ - \mu }^{\left(  +  \right)}  = \frac{{f_{ -  + } }}{{\sqrt 2 }},
\nonumber\\
&&
 h_{ + \mu }^{\left(  -  \right)}  = \frac{{f_{ +  - } }}{{\sqrt 2 }}, 
  \quad
h_{ - \mu }^{\left(  -  \right)}  = \frac{{f_{ -  - } }}{{\sqrt 2 }},
\label{Eq:mu-tau-h-2}
\end{eqnarray}
from the Yukawa interactions given by Eq.(\ref{Eq:FlavorCoulingPlusMinus-1})
for ($N_+$, $N_-$) and 
\begin{eqnarray}
&&
 h_{ + e}^{\left(  +  \right)}  = f_{ e e} , 
  \quad
h_{ - e}^{\left(  +  \right)}  = f_{ + e} , 
 \quad
h_{ + \mu }^{\left(  +  \right)}  = \frac{{f_{ e  + } }}{{\sqrt 2 }}, 
 \quad
h_{ - \mu }^{\left(  +  \right)}  = \frac{{f_{ +  + } }}{{\sqrt 2 }},
\nonumber\\
&&
 h_{ + \mu }^{\left(  -  \right)}  = \frac{{f_{ e  - } }}{{\sqrt 2 }}, 
  \quad
h_{ - \mu }^{\left(  -  \right)}  = \frac{{f_{ +  - } }}{{\sqrt 2 }},
\label{Eq:CouplingFlavorMasses2}
\end{eqnarray}
from
\begin{eqnarray}
&& 
-{\mathcal L}_\nu
=
\left( {f_{ee} {\overline {N_ e}}  + f_{ + e} {\overline {N_ +}} } \right)\nu _e  H_{u1}
+ \left( {f_{e + } {\overline {N_ e}}  + f_{ +  + } {\overline {N_ +}} } \right)\nu _ + H_{u1}
+ \left( {f_{ e  - } {\overline {N_ e}} + f_{+ - } {\overline {N_ +}}  } \right)\nu _ -  H_{u1},
\label{Eq:FlavorCoulingPlusMinus-2-e}
\end{eqnarray}
for $N=(N_e, N_+)$. Similarly for the other cases.  
Since $N$ can couple to $\nu_+$ but not to $\nu_-$ in the $\mu$-$\tau$ symmetric limit, flavor neutrino mass terms 
consist of $\nu_e\nu_+$ and $\nu_+\nu_+$, which are phenomenologically favorable \cite{NuPlus}.  Furthermore,
$U^ *{\bf Y}^{(+)}_\nu {\bf Y}^{(+)\dagger}_\nu U^T $  
becomes complex even in the $\mu$-$\tau$ symmetric limit due to the presence of 
the $\mu$-$\tau$ symmetric Majorana phase $\omega$ and may be 
preferable to the leptogenesis.
In the Yukawa interactions,
phases of the couplings of $f$'s can be absorbed into those associated with $\nu_{e,\pm}$.  However, in general 
we cannot make the $\mu$-$\tau$ symmetry preserving couplings real.  We have to adjust the $\nu_{e,+}$-couplings to 
be real by hand so that the $\mu$-$\tau$ symmetry breaking associated with the $\nu_-$-couplings supplies CP-phases.
\end{enumerate}
The major conclusion is that $N$ subject to the $\mu$-$\tau$ symmetry has real $\mu$-$\tau$ symmetry preserving couplings.  
Therefore, the leptonic CP-phases come from the $\mu$-$\tau$ symmetry breaking couplings. 
For $N$ blind to the $\mu$-$\tau$ symmetry, the same situation arises only if 
we assume that phases are associated with the $\nu_-$-couplings.  
In the $\mu$-$\tau$ symmetric limit,
the flavor neutrino masses can be parameterized to be:
\begin{eqnarray}
&&
\frac{1}{2}a_0 \nu _e \nu _e  + b_0 \frac{{\nu _e \nu _ +   + \nu _ +  \nu _e }}{{\sqrt 2 }} + d_ +  \nu _ +  \nu _ +   + d_ -  \nu _ -  \nu _ -  ,
\label{Eq:NuMassTermSym-1-0}
\end{eqnarray}
as in Eq.(\ref{Eq:NuMassTermSym}) for $N$ subject to the $\mu$-$\tau$ symmetry, and
\begin{eqnarray}
&&
\frac{1}{2}a_0 \nu _e \nu _e  + b_0 \frac{{\nu _e \nu _ +   + \nu _ +  \nu _e }}{{\sqrt 2 }} + d_ +  \nu _ +  \nu _ +,
\label{Eq:NuMassTermSym-3-0}
\end{eqnarray}
as in Eq.(\ref{Eq:NuMassTermSym-2}) for $N$ blind to the $\mu$-$\tau$ symmetry, where $a_0$. $b_0$ and $d_\pm$ are 
mass parameters.

\section{\label{sec:3}Flavor Neutrino Mass Matrix}
The seesaw mechanism generates the following Majorana neutrino mass matrix $M_\nu$ for flavor neutrinos:
\begin{eqnarray}
&& 
M_\nu   =  - v^2 {\bf Y}_\nu ^T U^T {\bf M}_{{\rm{diag}}}^{ - 1} U {\bf Y}_\nu,  
\label{Eq:SeesawMajorana}
\end{eqnarray}
where $v=\langle 0 \vert H_{u1}\vert 0\rangle$.
We estimate effects of the leptonic CP violation provided by $M_{ij}$ ($i,j=e,\mu,\tau$) defined 
in the Appendix \ref{sec:Appendix1}.  For the sake of simplicity, we assume that there is 
no phase in ${\bf M}_R$.\footnote{We keep terms containing the phase in our calculations, whose 
results are shown in the Appendix and will be used in our future study.}  

\subsection{\label{sec:3-1}$N$ subject to the $\mu$-$\tau$ Symmetry}
From Eq.(\ref{Eq:MassFlavorAppendix-1}), we find that 
\begin{eqnarray}
&&
 M_{ee}  \approx  - v^2 h_{ + e}^{\left(  +  \right)2} M_1^{ - 1},
\nonumber\\
&&
 M_{e\mu }^{\left(  +  \right)}  \approx  - v^2 h_{ + e}^{\left(  +  \right)} h_{ + \mu }^{\left(  +  \right)} M_1^{ - 1},
\nonumber\\
&&
 M_{e\mu }^{\left(  -  \right)}  \approx  - v^2 \left[ {\left( {h_{ + \mu }^{\left(  -  \right)}  - s h_{ - \mu }^{\left(  +  \right)} } \right)h_{ + e}^{\left(  +  \right)} M_1^{ - 1}  + \left( {h_{ - e}^{\left(  -  \right)}  + s h_{ + e}^{\left(  +  \right)} } \right)h_{ - \mu }^{\left(  +  \right)} M_2^{ - 1} } \right],
\nonumber\\
&&
 M_{\mu \mu }^{\left(  +  \right)}  \approx  - v^2 \left( {h_{ + \mu }^{\left(  +  \right)2} M_1^{ - 1}  + h_{ - \mu }^{\left(  +  \right)2} M_2^{ - 1} } \right),
\nonumber\\
&&
 M_{\mu \mu }^{\left(  -  \right)}  \approx  - 2v^2 \left[ {\left( {h_{ + \mu }^{\left(  -  \right)}  - s h_{ - \mu }^{\left(  +  \right)} } \right)h_{ + \mu }^{\left(  +  \right)} M_1^{ - 1}  + \left( {h_{ - \mu }^{\left(  -  \right)}  + s h_{ + \mu }^{\left(  +  \right)} } \right)h_{ - \mu }^{\left(  +  \right)} M_2^{ - 1} } \right],
\nonumber\\
&&
 M_{\mu \tau }  \approx  - v^2 \left( { - \sigma } \right)\left( {h_{ + \mu }^{\left(  +  \right)2} M_1^{ - 1}  - h_{ - \mu }^{\left(  +  \right)2} M_2^{ - 1} } \right),
\label{Eq:FlavorMasses}
\end{eqnarray}
up to the first order in the $\mu$-$\tau$ symmetry breaking terms $h_{- e}^{\left(  -  \right)}$, 
$h_{+ \mu }^{\left(  -  \right)}$, $h_{- \mu }^{\left(  -  \right)}$ and $s\propto{\bf M}_{+ -}$.
The results accord with our naive expectation that the $\mu$-$\tau$ symmetric masses of 
$M_{ee}$, $M^{(+)}_{e\mu}$ and $M^{(+)}_{\mu\mu}$ are controlled 
by the $\mu$-$\tau$ symmetric couplings. 
Since phases only arise from the $\mu$-$\tau$ symmetry breaking couplings, 
we readily observe that this property shows the followings:
\begin{enumerate}
\item the $\mu$-$\tau$ symmetric parts: $M_{ee}$, $M^{(+)}_{e\mu}$, $M^{(+)}_{\mu\mu}$ and $M_{\mu\tau}$ 
are real, and
\item the $\mu$-$\tau$ symmetry breaking parts: $M^{(-)}_{e\mu}$ and $M^{(-)}_{\mu\mu}$ are complex,
\end{enumerate}
within our approximation. It should be noted that $M_{ee}$, $M^{(+)}_{e\mu}$, $M^{(+)}_{\mu\mu}$ and $M_{\mu\tau}$
 would have imaginary parts if second order terms of the $\mu$-$\tau$ symmetry breaking are included.
For example, in $M_{ee}$, its phase arises from $h^{(-)}_{- e}$ because $\omega=0$. 
We then have a complex term proportional to $h^{(+)}_{+ e}h^{(-)}_{- e}$, whose coefficient 
is $cs$ which is the first order quantity because $s=0$ in 
the $\mu$-$\tau$ symmetric limit.  Since $h^{(-)}_{e\mu}$ being  
itself is the first order quantity, the whole contribution from this term is the 
second order quantity, which can be safely neglected.  In this way, we confirm that $M_{ee}$ can 
be almost real.  Similarly, we can confirm 
that $M^{(+)}$ itself is almost real. 
Therefore, main contributions to leptonic CP violation are given by $M^{(-)}_{e\mu}$ and $M^{(-)}_{\mu\mu}$. We will
specify the phases of $M^{(-)}_{e\mu}$ and $M^{(-)}_{\mu\mu}$ by $\alpha$ and $\beta$, respectively.

As suggested by the above phase structure, the flavor neutrino mass matrix can be 
parameterized by $M_\nu$: $M_\nu = M^{(+)}_\nu+M^{(-)}_\nu$ with
\begin{eqnarray}
&& 
M^{(+)}_\nu=\left( 
\begin{array}{*{20}c}
   a_0 & b_0 & - \sigma b_0 \\
   b_0 & d_0 & \sigma e_0  \\
   - \sigma b_0& \sigma e_0  & d_0\\
\end{array}
\right),
\quad
M^{(-)}_\nu=\left( 
\begin{array}{*{20}c}
   0 & b'_0 e^{i\alpha } & \sigma b'_0 e^{i\alpha } \\
   b'_0 e^{i\alpha } & d'_0 e^{i\beta } & 0  \\
   \sigma b'_0 e^{i\alpha } & 0  & - d'_0 e^{i\beta }  \\
\end{array}
\right),
\label{Eq:NuMatrixSimple}
\end{eqnarray}
where $a_0$, $b_0$, $d_0$, $e_0$, $b^\prime_0$ and $d^\prime_0$ are all real 
and $\alpha$ and $\beta$ are phases.  
We also use the notation: $a=a_0$, $b= b_0+b'_0 e^{i\alpha }$, 
$c= - \sigma \left(b_0-b'_0 e^{i\alpha }\right)$, 
$d= d_0+d'_0 e^{i\beta }$, $e= \sigma e_0$ and 
$f= d_0-d'_0 e^{i\beta }$. 
In approximately $\mu$-$\tau$ symmetric models, $b^\prime_0\approx 0$ and 
$d^\prime_0\approx 0$ are realized and yield the smallness of $\sin^2\theta_{13}$ and 
the large mixing given by $\sin 2\theta_{23}\approx 1$ as a natural consequence. 
However, to understand the observed smallness 
of $\Delta m^2_\odot/\vert\Delta m^2_{atm}\vert$, where $\Delta m^2_\odot = m^2_2-m^2_1(>0)$ and 
$\Delta m^2_{atm} = m^2_3-m^2_1$, needs an additional 
small parameter, which we call $\eta$.  

The $\mu$-$\tau$ symmetric mass matrix $M^{(+)}_\nu$ gives
\begin{eqnarray}
&&
\sin 2\theta _{12}  = \frac{{2\sqrt 2 b_0 }}{{\sqrt {\left( {d_0  - e_0  - a_0 } \right)^2  + 8b_0^2 } }},
\quad
\cos 2\theta_{23} = \sin\theta_{13} = 0,
\label{Eq:PredictionsMixingSym}\\
&&
m_1  = \frac{{a_0  + d_0  - e_0 }}{2} - \frac{{\sqrt 2 b_0 }}{{\sin 2\theta _{12} }},
\quad
m_2  = \frac{{a_0  + d_0  - e_0 }}{2} + \frac{{\sqrt 2 b_0 }}{{\sin 2\theta _{12} }},
\nonumber\\
&&
m_3  = d_0  + e_0,
\label{Eq:PredictionsMassSym}
\end{eqnarray}
as well as 
\begin{eqnarray}
&&
\Delta m^2_\odot = \frac{2\sqrt{2}({a_0  + d_0  - e_0 })b_0}{\sin 2\theta_{12}}.
\label{Eq:PredictionsAtmSolSym}
\end{eqnarray}
The spectrum contains one massless neutrino if the relation
\begin{eqnarray}
&&
a = \frac{{b\left( {bf - ce} \right) - c\left( {be - cd} \right)}}{{df - e^2 }},
\label{Eq:PredictionsMasslessSym1}
\end{eqnarray}
for $df \neq e^2$, is satisfied.  For $df= e^2$, we find that $be=cd$, leading to $m_3=0$ for $M^{(+)}_\nu$.  
Similarly, 
\begin{eqnarray}
&&
e = \frac{{bc \pm \sigma \sqrt {\left( {b^2  - ad} \right)\left( {c^2  - af} \right)} }}{a},
\label{Eq:PredictionsMasslessSym2}
\end{eqnarray}
is another useful relation.  
For $M^{(+)}_\nu$, it is readily understood that
\begin{enumerate}
\item for the normal mass hierarchy, $m_1=0$ is obtained 
form Eq.(\ref{Eq:PredictionsMasslessSym1}) if $a_0  + d_0  - e_0  \geq 0$,
\item for the inverted mass hierarchy, $m_3=0$ is obtained 
form Eq.(\ref{Eq:PredictionsMasslessSym2}) giving $d_0  + e_0=0$.
\end{enumerate}

The minimal seesaw mechanism in this case only allows the normal mass hierarchy \cite{mu-tauNoCP} 
to account for the observed results 
as discussed in the Appendix \ref{sec:Appendix1}.  One of the authors (M.Y.) has shown variety of textures, 
which are approximately $\mu$-$\tau$ symmetric \cite{NuPlus}, from which we choose 
the following neutrino mass matrix:
\begin{eqnarray}
&&
M^{(+)}_\nu
=
m_0 \left( {\begin{array}{*{20}c}
   {p\eta } & \eta& - \sigma \eta \\
   \eta & 1 & \sigma \left( {1 - s\eta } \right)  \\
  - \sigma \eta & {\sigma \left( {1 - s\eta } \right)} & 1 \\
\end{array}} \right)\quad (p=\frac{2}{s}),
\label{Eq:Normal-13-1}
\end{eqnarray}
giving
\begin{eqnarray}
&&
\tan 2\theta _{12} = \frac{2\sqrt{2}}{s-p},
\label{Eq:Normal-13-1-angles}
\end{eqnarray}
where $\eta$ is to be estimated in Sec.\ref{sec:4} to give 
$\eta (\sim\sqrt{\Delta m^2_\odot/\vert\Delta m^2_{atm}\vert})$=${\cal{O}}(10^{-1})$ and
$s$ (and $p$) are parameters of ${\cal {O}}(1)$.
The condition of $\det(M_\nu)=0$ is satisfied by 
\begin{eqnarray}
&&
a = \frac{{b\left( {bf - ce} \right) - c\left( {be - cd} \right)}}{{df - e^2 }}(=\frac{{2\eta }}{s}m_0~{\rm for}~M^{(+)}_\nu).
\label{Eq:Normal-13-1-zro}
\end{eqnarray}

The minimal seesaw model yields
\begin{eqnarray}
&&
 m_0  =  - v^2 \left( {h_{ + \mu }^{\left(  +  \right)2} M_1^{ - 1}  + h_{ - \mu }^{\left(  +  \right)2} M_2^{ - 1} } \right),
\nonumber\\
&&
 \eta  = \frac{{h_{ + e}^{\left(  +  \right)} h_{ + \mu }^{\left(  +  \right)} M_1^{ - 1} }}{{h_{ + \mu }^{\left(  +  \right)2} M_1^{ - 1}  + h_{ - \mu }^{\left(  +  \right)2} M_2^{ - 1} }},
\nonumber\\
&&
 s = 2\frac{{h_{ + \mu }^{\left(  +  \right)} }}{{h_{ + e}^{\left(  +  \right)} }},
\label{Eq:MassParamSeesaw}
\end{eqnarray}
where $p=2/s$ is automatically satisfied as expected.  Since $s={\cal{O}}(1)$ and $\eta={\cal{O}}(10^{-1})$ , we have to 
adjust the parameters such that
\begin{eqnarray}
&&
\left| {h_{ + \mu }^{\left(  +  \right)} } \right| \sim \left| {h_{ + e}^{\left(  +  \right)} } \right|
\ll 
\left| h_{ - \mu }^{\left(  +  \right)}\right|  \sqrt{\left| {M_1/M_2} \right|},
\label{Eq:MassParamSeesawAdjusted}
\end{eqnarray}
equivalently,
\begin{eqnarray}
&&
\left|f^{(+)}_{+ e}\right| \sim \left|f^{(+)}_{++}\right| \ll\left|f^{(+)}_{-- }\right| \sqrt{\left| {M_1/M_2} \right|},
\label{Eq:MassParamSeesawAdjusted1}
\end{eqnarray}
which gives $\nu_-\nu_-$ as a dominant mass term.

\subsection{\label{sec:3-2}$N$ blind to the $\mu$-$\tau$ Symmetry}
From Eq.(\ref{Eq:MassFlavorAppendix-2}), we find that 
\begin{eqnarray}
&&
 M_{ee}  =  - v^2 \left[ {\left( {ch_{ + e}^{\left(  +  \right)}  - s h_{ - e}^{\left(  +  \right)} } \right)^2 M_1^{ - 1}  + \left( {s h_{ + e}^{\left(  +  \right)}  + ch_{ - e}^{\left(  +  \right)} } \right)^2 M_2^{ - 1} } \right],
\nonumber\\ 
&&
 M_{e\mu }^{\left(  +  \right)}  =  - v^2 \left[ {\left( {ch_{ + e}^{\left(  +  \right)}  - s h_{ - e}^{\left(  +  \right)} } \right)\left( {ch_{ + \mu }^{\left(  +  \right)}  - s h_{ - \mu }^{\left(  +  \right)} } \right)M_1^{ - 1}  + \left( {s h_{ + e}^{\left(  +  \right)}  + ch_{ - e}^{\left(  +  \right)} } \right)\left( {s h_{ + \mu }^{\left(  +  \right)}  + ch_{ - \mu }^{\left(  +  \right)} } \right)M_2^{ - 1} } \right],
\nonumber\\ 
&&
 M_{e\mu }^{\left(  -  \right)}  =  - v^2 \left[ {\left( {ch_{ + e}^{\left(  +  \right)}  - s h_{ - e}^{\left(  +  \right)} } \right)\left( {ch_{ + \mu }^{\left(  -  \right)}  - s h_{ - \mu }^{\left(  -  \right)} } \right)M_1^{ - 1}  + \left( {s h_{ + e}^{\left(  +  \right)}  + ch_{ - e}^{\left(  +  \right)} } \right)\left( {s h_{ + \mu }^{\left(  -  \right)}  + ch_{ - \mu }^{\left(  -  \right)} } \right)M_2^{ - 1} } \right],
\nonumber\\ 
&&
 M_{\mu \mu }^{\left(  +  \right)}  \approx  - v^2 \left[ {\left( {ch_{ + \mu }^{\left(  +  \right)}  - s h_{ - \mu }^{\left(  +  \right)} } \right)^2 M_1^{ - 1}  + \left( {ch_{ - \mu }^{\left(  +  \right)}  + s h_{ + \mu }^{\left(  +  \right)} } \right)^2 M_2^{ - 1} } \right],
\nonumber\\ 
&&
 M_{\mu \mu }^{\left(  -  \right)} {\rm{ }} =  - 2v^2 \left[ {\left( {ch_{ + \mu }^{\left(  +  \right)}  - s h_{ - \mu }^{\left(  +  \right)} } \right)\left( {ch_{ + \mu }^{\left(  -  \right)}  - s h_{ - \mu }^{\left(  -  \right)} } \right)M_1^{ - 1}  + \left( {ch_{ - \mu }^{\left(  +  \right)}  + s h_{ + \mu }^{\left(  +  \right)} } \right)\left( {ch_{ - \mu }^{\left(  -  \right)}  + s h_{ + \mu }^{\left(  -  \right)} } \right)M_2^{ - 1} } \right],
\nonumber\\ 
&&
 M_{\mu \tau }  \approx  - v^2 \left( { - \sigma } \right)\left[ {\left( {ch_{ + \mu }^{\left(  +  \right)}  - s h_{ - \mu }^{\left(  +  \right)} } \right)^2 M_1^{ - 1}  + \left( {ch_{ - \mu }^{\left(  +  \right)}  + s h_{ + \mu }^{\left(  +  \right)} } \right)^2 M_2^{ - 1} } \right],
\label{Eq:FlavorMasses2}
\end{eqnarray}
up to the first order in the $\mu$-$\tau$ symmetry breaking terms $h_{+ \mu}^{\left(  -  \right)}$, 
and $h_{- \mu }^{\left(  -  \right)}$.

The minimal seesaw mechanism forbids the normal mass hierarchy to account for the observed results 
as discussed in the Appendix \ref{sec:Appendix1}.  There are two types of neutrino mass textures \cite{NuPlus}.
\begin{enumerate}
\item As the inverted mass hierarchy I (with $m_1\sim m_2$),
\begin{eqnarray}
&&
M^{(+)}_\nu
=
m_0 \left( {\begin{array}{*{20}c}
   {2-p\eta } & \eta& -\sigma \eta \\
   \eta & 1 & -\sigma \\
  - \sigma \eta & -\sigma & 1 \\
\end{array}} \right),
\label{Eq:Inverted-13-1}\\
\end{eqnarray}
leading to
\begin{eqnarray}
&&
\tan 2\theta _{12} = \frac{2\sqrt{2}}{p}.
\label{Eq:Inverted-13-1-angles}
\end{eqnarray}
The condition of $\det(M_\nu)=0$ is satisfied by
\begin{eqnarray}
&&
e = \frac{{bc - \sigma \sqrt {\left( {b^2  - ad} \right)\left( {c^2  - af} \right)} }}{a}
(=-\sigma d_0~{\rm for}~M^{(+)}_\nu).
\label{Eq:Inverted-13-1-zro}
\end{eqnarray}
These parameters are related to those in the seesaw mechanism given by
\begin{eqnarray}
&&
 m_0  =  - v^2 \left[ {\left( {ch_{ + \mu }^{\left(  +  \right)}  - s h_{ - \mu }^{\left(  +  \right)} } \right)^2 M_1^{ - 1}  + \left( {ch_{ - \mu }^{\left(  +  \right)}  + s h_{ + \mu }^{\left(  +  \right)} } \right)^2 M_2^{ - 1} } \right],
\nonumber\\
&&
 \eta  = \frac{{\left( {ch_{ + e}^{\left(  +  \right)}  - s h_{ - e}^{\left(  +  \right)} } \right)\left( {ch_{ + \mu }^{\left(  +  \right)}  - s h_{ - \mu }^{\left(  +  \right)} } \right)M_1^{ - 1}  + \left( {s h_{ + e}^{\left(  +  \right)}  + ch_{ - e}^{\left(  +  \right)} } \right)\left( {s h_{ + \mu }^{\left(  +  \right)}  + ch_{ - \mu }^{\left(  +  \right)} } \right)M_2^{ - 1} }}{{\left( {ch_{ + \mu }^{\left(  +  \right)}  - s h_{ - \mu }^{\left(  +  \right)} } \right)^2 M_1^{ - 1}  + \left( {ch_{ - \mu }^{\left(  +  \right)}  + s h_{ + \mu }^{\left(  +  \right)} } \right)^2 M_2^{ - 1} }},
\nonumber\\
&&
p = \frac{{\left[ {2\left( {ch_{ + \mu }^{\left(  +  \right)}  - s h_{ - \mu }^{\left(  +  \right)} } \right)^2  - \left( {ch_{ + e}^{\left(  +  \right)}  - s h_{ - e}^{\left(  +  \right)} } \right)^2 } \right]M_1^{ - 1}  + \left[ {2\left( {ch_{ - \mu }^{\left(  +  \right)}  + s h_{ + \mu }^{\left(  +  \right)} } \right)^2  - \left( {s h_{ + e}^{\left(  +  \right)}  + ch_{ - e}^{\left(  +  \right)} } \right)^2 } \right]M_2^{ - 1} }}{{\left( {ch_{ + e}^{\left(  +  \right)}  - s h_{ - e}^{\left(  +  \right)} } \right)\left( {ch_{ + \mu }^{\left(  +  \right)}  - s h_{ - \mu }^{\left(  +  \right)} } \right)M_1^{ - 1}  + \left( {s h_{ + e}^{\left(  +  \right)}  + ch_{ - e}^{\left(  +  \right)} } \right)\left( {s h_{ + \mu }^{\left(  +  \right)}  + ch_{ - \mu }^{\left(  +  \right)} } \right)M_2^{ - 1} }},
\label{Eq:MassParamSeesaw2-1}
\end{eqnarray}
\item As the inverted mass hierarchy I\hspace{-.1em}I (with $m_1\sim -m_2$),
\begin{eqnarray}
&&
M^{(+)}_\nu
=
m_0 \left( {\begin{array}{*{20}c}
   -\left(2-\eta\right) & q& -\sigma q \\
   q & 1 & -\sigma \\
  - \sigma q & -\sigma & 1 \\
\end{array}} \right),
\label{Eq:Inverted-13-2}
\end{eqnarray}
leading to
\begin{eqnarray}
&&
\tan 2\theta _{12} = \frac{{2\sqrt 2 q}}{{4 - \eta }}.
\label{Eq:Inverted-13-2-angles}
\end{eqnarray}
The condition of $\det(M_\nu)=0$ is satisfied by
\begin{eqnarray}
&&
e = \frac{{bc + \sigma \sqrt {\left( {b^2  - ad} \right)\left( {c^2  - af} \right)} }}{a}
(=-\sigma d_0~{\rm for}~M^{(+)}_\nu).
\label{Eq:Inverted-13-2-zro}
\end{eqnarray}
These parameters are related to those in the seesaw mechanism given by
\begin{eqnarray}
&&
 m_0  =  - v^2 \left[ {\left( {ch_{ + \mu }^{\left(  +  \right)}  - s h_{ - \mu }^{\left(  +  \right)} } \right)^2 M_1^{ - 1}  + \left( {ch_{ - \mu }^{\left(  +  \right)}  + s h_{ + \mu }^{\left(  +  \right)} } \right)^2 M_2^{ - 1} } \right],
\nonumber\\
&&
 \eta  = \frac{{\left[ {2\left( {ch_{ + \mu }^{\left(  +  \right)}  - s h_{ - \mu }^{\left(  +  \right)} } \right)^2  + \left( {ch_{ + e}^{\left(  +  \right)}  - s h_{ - e}^{\left(  +  \right)} } \right)^2 } \right]M_1^{ - 1}  + \left[ {2\left( {ch_{ - \mu }^{\left(  +  \right)}  + s h_{ + \mu }^{\left(  +  \right)} } \right)^2  + \left( {s h_{ + e}^{\left(  +  \right)}  + ch_{ - e}^{\left(  +  \right)} } \right)^2 } \right] M_2^{ - 1} }}{{\left( {ch_{ + \mu }^{\left(  +  \right)}  - s h_{ - \mu }^{\left(  +  \right)} } \right)^2 M_1^{ - 1}  + \left( {ch_{ - \mu }^{\left(  +  \right)}  + s h_{ + \mu }^{\left(  +  \right)} } \right)^2 M_2^{ - 1} }},
\nonumber\\
&& q = \frac{{\left( {ch_{ + e}^{\left(  +  \right)}  - s h_{ - e}^{\left(  +  \right)} } \right)\left( {ch_{ + \mu }^{\left(  +  \right)}  - s h_{ - \mu }^{\left(  +  \right)} } \right)M_1^{ - 1}  + \left( {s h_{ + e}^{\left(  +  \right)}  + ch_{ - e}^{\left(  +  \right)} } \right)\left( {s h_{ + \mu }^{\left(  +  \right)}  + ch_{ - \mu }^{\left(  +  \right)} } \right)M_2^{ - 1} }}{{\left( {ch_{ + \mu }^{\left(  +  \right)}  - s h_{ - \mu }^{\left(  +  \right)} } \right)^2 M_1^{ - 1}  + \left( {ch_{ - \mu }^{\left(  +  \right)}  + s h_{ + \mu }^{\left(  +  \right)} } \right)^2 M_2^{ - 1} }},
\label{Eq:MassParamSeesaw2-2}
\end{eqnarray}
\end{enumerate}
The parameter $\eta$ is to be estimated in Sec.\ref{sec:4} to give 
$\eta (\sim\Delta m^2_\odot/\vert\Delta m^2_{atm}\vert)$=${\cal{O}}(10^{-2})$ and
$p$ and $q$ are parameters of ${\cal {O}}(1)$.  We have to adjust sizes of the parameters of 
the seesaw to account for the neutrino mass spectrum.

\section{\label{sec:4}CP Phases}
In this section, we discuss how leptonic CP phases are generated by $M_\nu$ of Eq.(\ref{Eq:NuMatrixSimple}).
For $N$ blind to the $\mu$-$\tau$ symmetry, the phase structure of Eq.(\ref{Eq:NuMatrixSimple}) is not a general 
consequence.  So, we choose a specific parameter set so that phases only arise from 
$M^{(-)}_{e\mu}$ and $M^{(-)}_{\mu\mu}$.  In other words, phases should be associated with the couplings of $\nu_-$.  
Furthermore, there have been arguments that the renormalization effects 
are significant for the inverted mass hierarchy \cite{RGE}, which is the case of $N$ blind to the $\mu$-$\tau$ symmetry.  
However, the smallness of $\sin^2\theta_{13}$ is not disturbed because it is a result of the approximate $\mu$-$\tau$ 
symmetry but the CP-violating phases may receive significant distortion. 
This subject will be discussed elsewhere. For a moment, 
we show the case of the inverted mass hierarchy to 
make a comparison with the case of the normal mass hierarchy.

Our seesaw model has four phases from three Yukawa couplings and one Majorana 
phase of heavy neutrinos corresponding to one Dirac phase and three Majorana phase where one 
overall Majorana phase is redundant.  Therefore, three CP-violating phases are present. 
This number is consistent with the general result of the seesaw model 
with $N$-flavor and $M$-heavy neutrinos, giving $N(M-1)$.  Since the $\mu$-$\tau$ 
symmetry breaking is so small that terms up to its first order contributions as in 
Eq.(\ref{Eq:FlavorMasses}) can 
well describe neutrino phenomenology, two phases $\alpha$ and $\beta$ become active 
and other phases associated with second-order contributions are safely neglected.  
The CP-phases including $\delta_{CP}=\delta+\rho$
 are in general complicated functions of $\alpha$ and $\beta$. 
 These two phases are the sources of the Dirac and Majorana phases in $U_{PMNS}$.
 However, we will see that 
 when mass 
hierarchies are taken into account, $\rho$ is found to be small and the 
dependence of $\alpha$ and $\beta$ can be derived to give $\delta_{CP} \sim \alpha$
 for the normal mass hierarchy and $\delta_{CP} \sim -\alpha$ the inverted mass 
hierarchy (with $m_1 \sim m_2$).  These features can be viewed in the figures of 
$\delta_{CP}$ to be presented.

\subsection{\label{sec:4-1}Estimations}
The Dirac CP-violating phase is given by $\delta+\rho$ evaluated from Eq.(\ref{Eq:Mdagger-ExactMixingAngles23})
in the Appendix \ref{sec:Appendix2}, from which we obtain that
\begin{eqnarray}
&&
c_{13} X \approx \sqrt 2 \left( { b_0 \left( {a_0 + d_0  - e_0 } \right) + b'_0 d'_0 e^{i\left( {\beta  - \alpha } \right)}  + \left( {\Delta  + i\gamma } \right)\left( {a_0 b'_0 e^{i\alpha }  + b'_0 e^{ - i\alpha } \left( {d_0  + e_0 } \right) + b_0 d'_0 e^{i\beta } } \right)} \right),
\label{Eq:X}\\
&&
Y \approx \sqrt 2 \sigma \left( - {\left( { \Delta - i\gamma } \right)\left( { b_0 \left( {a_0 + d_0  - e_0 } \right) + b'_0 d'_0 e^{i\left( {\beta  - \alpha } \right)} } \right) + a_0 b'_0 e^{i\alpha }  + b'_0 e^{ - i\alpha } \left( {d_0  + e_0 } \right) + b_0 d'_0 e^{i\beta } } \right),
\label{Eq:Y}
\end{eqnarray}
where the approximation is due to $\vert \gamma\vert \ll 1$, 
$\cos\theta_{23}=(1+\Delta)/\sqrt{2}$ and $\sin\theta_{23}= \sigma(1-\Delta)/\sqrt{2}$ for 
$\vert \Delta\vert \ll 1$.  The phases $\delta$ and $\rho$ are calculated from
\begin{eqnarray}
&&
\delta = -\arg(Y), 
\quad
\rho = \arg(X).
\label{Eq:X-Y}
\end{eqnarray}
From Eq.(\ref{Eq:X}), it is 
expected that $\rho\approx 0$ if $b_0 \left( {a_0 + d_0  - e_0} \right)$ is not suppressed.  This expectation 
is valid in the two textures of the inverted mass hierarchy; however, $\rho$ may not be suppressed 
in the normal mass hierarchy because $a_0 + d_0  - e_0\approx 0$ by Eq.(\ref{Eq:Normal-13-1}).  
The parameters $\gamma$ and $\Delta$ are estimate to be:
\begin{eqnarray}
&&
\gamma \approx \frac{{4\left( {b_0 b'_0 \sin \alpha  - e_0 d'_0 \sin \beta } \right) - \sigma \sin\theta_{13}\sin 2\theta _{12} \sin \left( {\rho  + \delta } \right) \Delta m_ \odot ^2 }}{{2\Delta m_{atm}^2 }},
\label{Eq:gamma}\\
&&
\Delta  \approx  - \frac{{4\left( {b_0 b'_0 \cos \alpha  + d_0 d'_0 \cos \beta } \right) + \sigma\sin\theta_{13}\sin 2\theta _{12} \cos \left( {\rho  + \delta } \right) \Delta m_ \odot ^2 }}{{2\Delta m_{atm}^2 }}.
\label{Eq:Delta}
\end{eqnarray}
The CP-violating phase $\delta+ \rho$ can be numerically obtained from Eqs.(\ref{Eq:X}) and (\ref{Eq:Y}) by using iteration, 
where $\Delta \pm i\gamma$ is given by
$
\gamma \approx  {2 \left( {b_0 b'_0 \sin \alpha  - e_0 d'_0 \sin \beta } \right)}/{\Delta m_{atm}^2 }
$
and
$
\Delta  \approx  - {2\left( {b_0 b'_0 \cos \alpha  + d_0 d'_0 \cos \beta } \right)}/{\Delta m_{atm}^2 }
$
as a first trial.

The CP-violating Majorana phase is estimated from Eq.(\ref{Eq:M}) for $m_{1,2,3}$.  
We have assured, as expected, that $m_1= 0$ for the normal mass hierarchy and 
$m_3= 0$ for the inverted mass hierarchy within our numerical accuracy.
From Eq.(\ref{Eq:M}), we find that
\begin{eqnarray}
&&
 m_2 e^{ - 2i\phi_2 }  \approx \frac{{2\sqrt 2 }}{{\sin 2\theta _{12} }}\left[ {\left( {1 + i\gamma \Delta } \right)b_0  + \left( {\Delta  + i\gamma } \right)b'_0 e^{i\alpha } } \right]e^{i\rho },
\nonumber\\
&&
 m_3 e^{ - 2i\phi_3 }  \approx \lambda _3  + s_{13}^2 \left( {\lambda _3  - e^{2i\delta } a} \right), 
\label{Eq:normal-Majorana-phase}
\end{eqnarray}
where $\lambda _3  \approx d_0  + e_0  - 2i\left( {2\Delta \gamma d_0  - \left( {\gamma  + i\Delta } \right)e^{i\beta } d'_0 } \right)$, 
for the normal mass hierarchy with $m_1=0$, and 
\begin{eqnarray}
 m_1 e^{ - 2i\phi_1 }  &\approx& \frac{{ae^{2i\rho }  + d_0  - e_0  + 2\left( {2i\Delta \gamma d_0  + \left( {\Delta  + i\gamma } \right)e^{i\beta } d'_0 } \right)}}{2} 
\nonumber\\
&&
 - \frac{{\sqrt 2 }}{{\sin 2\theta _{12} }}\left[ {\left( {1 + i\gamma \Delta } \right)b_0  + \left( {\Delta  + i\gamma } \right)b'_0 e^{i\alpha } } \right]e^{i\rho },
\nonumber
\end{eqnarray}
\begin{eqnarray}
 m_2 e^{ - 2i\phi_2 }  &\approx& \frac{{ae^{2i\rho }  + d_0  - e_0  + 2\left( {2i\Delta \gamma d_0  + \left( {\Delta  + i\gamma } \right)e^{i\beta } d'_0 } \right)}}{2} 
\nonumber\\
&&
+ \frac{{\sqrt 2 }}{{\sin 2\theta _{12} }}\left[ {\left( {1 + i\gamma \Delta } \right)b_0  + \left( {\Delta  + i\gamma } \right)b'_0 e^{i\alpha } } \right]e^{i\rho },
\label{Eq:inverted-Majorana-phase}
\end{eqnarray}
for the inverted mass hierarchy with $m_3=0$.
It should be noted that the size of $\phi_{1,2}$ is generically small
since the nonvanishing $m_{1,2}$  for the inverted mass hierarchy start with 
the unsuppressed $\mu$-$\tau$ symmetric terms.  

To perform our numerical calculations, we use exact formula without approximation: 
Eq.(\ref{Eq:Mdagger-ExactMixingAngles23}) for $\theta_{12,13}$, $\delta$ and $\rho$, 
Eq.(\ref{Eq:Exact-cos23-solution}) for $\theta_{23}$
 Eq.(\ref{Eq:Exact-gamma-solution}) for $\gamma$  and 
Eq.(\ref{Eq:M}) for $\phi_{1,2,3}$.
The tri-bimaximal neutrino mixing \cite{TriBiMaximal} is assumed for 
$M^{(+)}_\nu$ and is realized by 
\begin{enumerate}
\item $s=2$ in Eq.(\ref{Eq:Normal-13-1}) for the normal mass hierarchy,
\item $p=1$ in Eq.(\ref{Eq:Inverted-13-1}) for the inverted mass hierarchy I (with $m_1\sim m_2$),
\item $q=4-\eta$ in Eq.(\ref{Eq:Inverted-13-2}) for the inverted mass hierarchy I\hspace{-.1em}I (with $m_1\sim -m_2$).
\end{enumerate}
We estimate the CP-violating phases $\delta+\rho$ and $\phi_1-\phi_2$ (or $\phi_2-\phi_3$)
as well as the mixing angles as functions of $\alpha$ and $\beta$ for given values of 
$\vert\Delta m^2_{atm}\vert= 2.59\sim 2.61$ $(\times 10^{-3}$ eV$^2$) and 
$\Delta m^2_\odot= 7.87\sim 7.93$ $(\times 10^{-5}$ eV$^2$), which are taken to sit on 
values around their center values 
in the recent data:
\begin{eqnarray}
&&
\vert\Delta m^2_{atm}\vert = \left(2.6\pm0.2\right) \times 10^{-3}~{\rm eV}^2,
\quad
\Delta m^2_\odot = \left(7.9\pm 0.3\right) \times 10^{-5}~{\rm eV}^2,
\label{Eq:NuData}
\end{eqnarray}
as the allowed 1$\sigma$ ranges \cite{NeutrinoData}.  Our iteration starts with the calculation of $m_0$ by using 
$D_+$ in Eq.(\ref{Eq:Mdagger-M-MassRelation}) for given values of 
$\Delta m^2_{atm}$ and $\Delta m^2_\odot$ and 
these given values are compared with their computed values from our formula for a consistency check.

\subsection{\label{sec:4-2}Predictions}
Before performing numerical calculations, we show our predictions from three textures:
\begin{enumerate}
\item for the normal mass hierarchy ($p=2/s$ with $s=2$),
\begin{eqnarray}
&&
\frac{\Delta m^2_\odot}{\vert\Delta m^2_{atm}\vert} \approx \frac{\left(s+p\right)\eta^2}{\sqrt{2}\sin 2\theta_{12}},
\label{Eq:Normal-eta}
\end{eqnarray}
suggesting that $\eta$=${\cal{O}}(10^{-1})$, and
\begin{eqnarray}
&&
c_{13} X \approx \sqrt  2 \left(  m^2_0{\left( {s + p} \right)\eta ^2 + b'_0 d'_0 e^{i\left( {\beta  - \alpha } \right)}  + 2m_0\left( {\Delta  + i\gamma } \right)b'_0 e^{ - i\alpha } }\right),
\quad
Y \approx \sqrt 2\sigma m_0b'_0 e^{ - i\alpha } ,
\label{Eq:Normal-XY}
\end{eqnarray}
leading to
\begin{eqnarray}
&&
\rho = {\rm arbitrary},
\quad
\delta \approx \alpha,
\label{Eq:NormalPhases}
\end{eqnarray}
where we will numerically find that the term proportional to $\eta^2$ gives dominated contribution in 
 $c_{13} X$, which result in $\rho \approx 0$, and
\begin{eqnarray}
&&
\sin\theta_{13}  e^{ - i\delta } \approx \frac{Y}{{\Delta m_{atm}^2 }},
\nonumber \\
&&
\gamma  \approx 2\left( { \eta b'_0\sin \alpha   - d'_0 \sin \beta } \right)\frac{{m_0 }}{{\Delta m_{atm}^2 }} - \frac{1}{2}\sigma s_{13} \sin \left( {\rho  + \delta } \right)\sin 2\theta _{12} \frac{{\Delta m_ \odot ^2 }}{{\Delta m_{atm}^2 }},
\nonumber \\
&&
\cos 2\theta _{23} \left(\approx 2\Delta\right) \approx  - \left( {4\left( { \eta b'_0\cos \alpha  + d'_0 \cos \beta } \right)\frac{{m_0 }}{{\Delta m_{atm}^2 }} +\sigma s_{13} \cos \left( {\rho  + \delta } \right)\sin 2\theta _{12} \frac{{\Delta m_ \odot ^2 }}{{\Delta m_{atm}^2 }}} \right),
\label{Eq:NormalPredictions}
\end{eqnarray}
as well as
\begin{eqnarray}
&&
 m_2 e^{ - 2i\phi _2 }  \approx \frac{{2\sqrt 2 \eta m_0 e^{i\rho } }}{{\sin 2\theta _{12} }},
\quad
 m_3 e^{ - 2i\phi _3 }  \approx 
\left( {2 - s\eta } \right)m_0  - 2\left( {2i\Delta \gamma m_0  + \left( {\Delta  - i\gamma } \right)e^{i\beta } d'_0 } \right),
\label{Eq:NormalPredictionsMasses}
\end{eqnarray}
leading to $\vert\Delta m^2_{atm}\vert \approx 4m^2_0$ and
\begin{eqnarray}
&&
{\phi  \approx  - \frac{{\rho}}{4}},
\label{Eq:Normal-phi}
\end{eqnarray}
for $\rho\approx 0$,
\item for the inverted mass hierarchy I (with $m_1\sim m_2$ with $p=1$),
\begin{eqnarray}
&&
\frac{\Delta m^2_\odot}{\vert\Delta m^2_{atm}\vert} \approx \frac{2\sqrt{2}\eta}{\sin 2\theta_{12}},
\label{Eq:Inverted1-eta}
\end{eqnarray}
suggesting that $\eta$=${\cal{O}}(10^{-2})$, and
\begin{eqnarray}
&&
 c_{13} X \approx 4\sqrt 2 m^2_0 \eta ,
\quad
 Y \approx 2\sqrt 2 \sigma m_0b'_0 e^{ i\alpha } ,
\label{Eq:Inverted1-XY}
\end{eqnarray}
leading to 
\begin{eqnarray}
&&
\rho \approx 0,
\quad
\delta \approx -\alpha,
\label{Eq:Inverted1Phases}
\end{eqnarray}
and
\begin{eqnarray}
&&
\sin\theta_{13} e^{ - i\delta }  \approx \frac{Y}{{\Delta m_{atm}^2 }},
\nonumber \\
&&
\gamma  \approx 2\left( {\eta b'_0 \sin \alpha  + d'_0 \sin \beta } \right)\frac{{m_0 }}{{\Delta m_{atm}^2 }} - \frac{1}{2}\sigma s_{13} \sin \left( {\rho  + \delta } \right)\sin 2\theta _{12} \frac{{\Delta m_ \odot ^2 }}{{\Delta m_{atm}^2 }},
\nonumber \\
&&
\cos 2\theta _{23} \left(\approx 2\Delta\right) \approx  
 - \left( {4\left( {\eta b'_0 \cos \alpha  + d'_0 \cos \beta } \right)\frac{{m_0 }}{{\Delta m_{atm}^2 }} + \sigma s_{13} \cos \left( {\rho  + \delta } \right)\sin 2\theta _{12} \frac{{\Delta m_ \odot ^2 }}{{\Delta m_{atm}^2 }}} \right),
\label{Eq:Inverted1Predictions}
\end{eqnarray}
as well as
\begin{eqnarray}
&&
 m_1 e^{ - 2i\phi_1 }  \approx \left( {1 + e^{2i\rho }  - \frac{{p\eta e^{2i\rho } }}{2} - \frac{{\sqrt 2 \eta e^{i\rho } }}{{\sin 2\theta _{12} }}} \right)m_0,
\nonumber \\
&&
 m_2 e^{ - 2i\phi_2 }  \approx \left( {1 + e^{2i\rho }  - \frac{{p\eta e^{2i\rho } }}{2} + \frac{{\sqrt 2 \eta e^{i\rho } }}{{\sin 2\theta _{12} }}} \right)m_0,
\label{Eq:Inverted1PredictionsMasses}
\end{eqnarray}
leading to $\vert\Delta m^2_{atm}\vert \approx m^2_0$ and
\begin{eqnarray}
&&
\phi = 0,
\label{Eq:inverted-phi}
\end{eqnarray}
up to $\mathcal{O}(\rho^2)$ and
\item for the inverted mass hierarchy I\hspace{-.1em}I (with $m_1\sim -m_2$ with $q=4-\eta$),
\begin{eqnarray}
&&
\frac{\Delta m^2_\odot}{\vert\Delta m^2_{atm}\vert} \approx \frac{\sqrt{2}\eta\sin 2\theta_{12}}{q},
\label{Eq:Inverted2-eta}
\end{eqnarray}
suggesting that $\eta$=${\cal{O}}(10^{-2})$, and
\begin{eqnarray}
&&
 c_{13} X \approx 2\sqrt 2 m^2_0\eta q,
\quad
 Y \approx  - \sqrt 2 \sigma m_0\left( {2b'_0 e^{i\alpha }  - qd'_0 e^{i\beta } } \right),
\label{Eq:Inverted2-XY}
\end{eqnarray}
leading to 
\begin{eqnarray}
&&
\rho \approx 0,
\quad
\delta \approx {\rm arbitrary},
\label{Eq:Inverted2Phases}
\end{eqnarray}
and
\begin{eqnarray}
&&
\sin\theta_{13} e^{ - i\delta }  \approx \frac{Y}{{\Delta m_{atm}^2 }},
\nonumber \\
&&
\gamma  \approx 2 \left( {qb'_0 \sin \alpha  + d'_0 \sin \beta } \right)\frac{{m_0 }}{{\Delta m_{atm}^2 }} - \frac{1}{2}\sigma s_{13} \sin \left( {\rho  + \delta } \right)\sin 2\theta _{12} \frac{{\Delta m_ \odot ^2 }}{{\Delta m_{atm}^2 }},
\nonumber \\
&&
\cos 2\theta _{23} \left(\approx 2\Delta\right) \approx  
 - 4\left( {qb'_0 \cos \alpha  + d'_0 \cos \beta } \right)\frac{{m_0 }}{{\Delta m_{atm}^2 }} + \sigma s_{13} \cos \left( {\rho  + \delta } \right)\sin 2\theta _{12} \frac{{\Delta m_ \odot ^2 }}{{\Delta m_{atm}^2 }},
\label{Eq:Inverted2Predictions}
\end{eqnarray}
as well as
\begin{eqnarray}
&&
m_1 e^{ - 2i\phi_1}  \approx \left( {\frac{{2\left( {1 - e^{2i\rho } } \right) + \eta e^{2i\rho } }}{2} - \frac{{\sqrt 2 qe^{i\rho } }}{{\sin 2\theta _{12} }}} \right)m_0, 
\nonumber \\
&&
m_2 e^{ - 2i\phi _2 }  \approx \left( {\frac{{2\left( {1 - e^{2i\rho } } \right) + \eta e^{2i\rho } }}{2} + \frac{{\sqrt 2 qe^{i\rho } }}{{\sin 2\theta _{12} }}} \right)m_0, 
\label{Eq:Inverted2PredictionsMasses}
\end{eqnarray}
leading to $\vert\Delta m^2_{atm}\vert \approx m^2_0$ and
\begin{eqnarray}
&&
\phi  \approx  - \frac{{\sin 2\theta _{12} }}{{\sqrt 2 q}}\rho,
\label{Eq:Inverted2-phi}
\end{eqnarray}
which becomes $- \rho/6$ for $\sin 2\theta_{12}\approx 2\sqrt{2}/3$ and $q\approx 4$.
\end{enumerate}
It is expected that $\sin\theta_{13}$ has no distinct dependence 
of $\alpha$ and $\beta$ because $\sin\theta_{13}$ is determined by the radial part 
of $Y$ whose phase from $\alpha$ and $\beta$ controls $\delta$.

\begin{figure}[!htbp]
\begin{flushleft}
\includegraphics*[20mm,162mm][195mm,263mm]{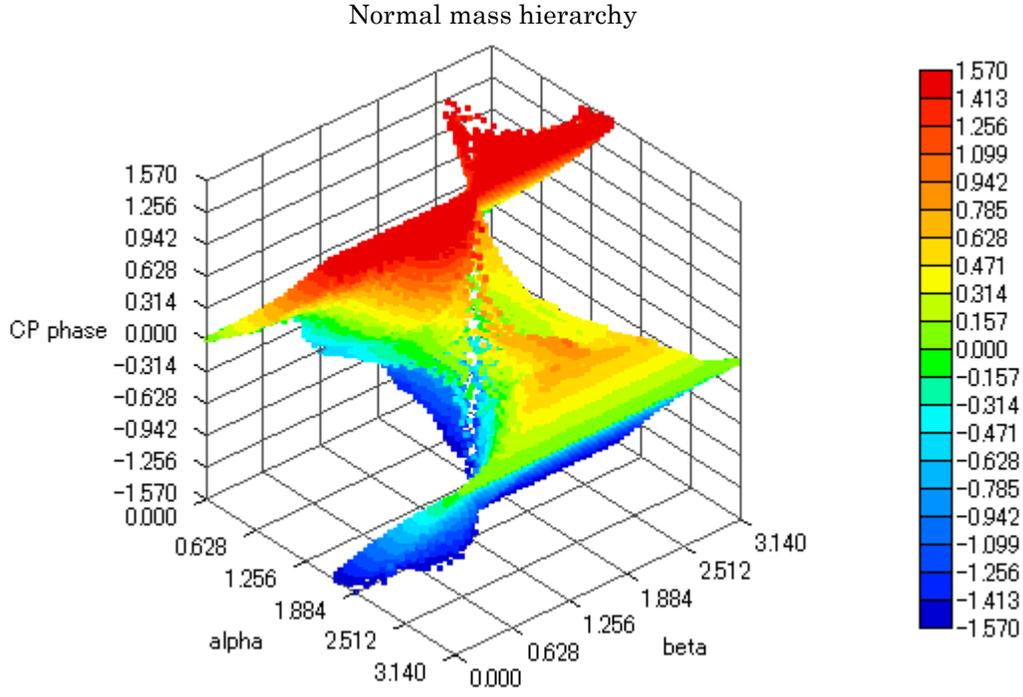}
\end{flushleft}
\vspace{-4mm}
\caption{The predictions of the Dirac CP phase $\delta+\rho$ as function of $\alpha$ and $\beta$ for the normal mass hierarchy.}
\label{Fig:phase-normal}
\end{figure}

\begin{figure}[!htbp]
\begin{flushleft}
\includegraphics*[13mm,161mm][205mm,263mm]{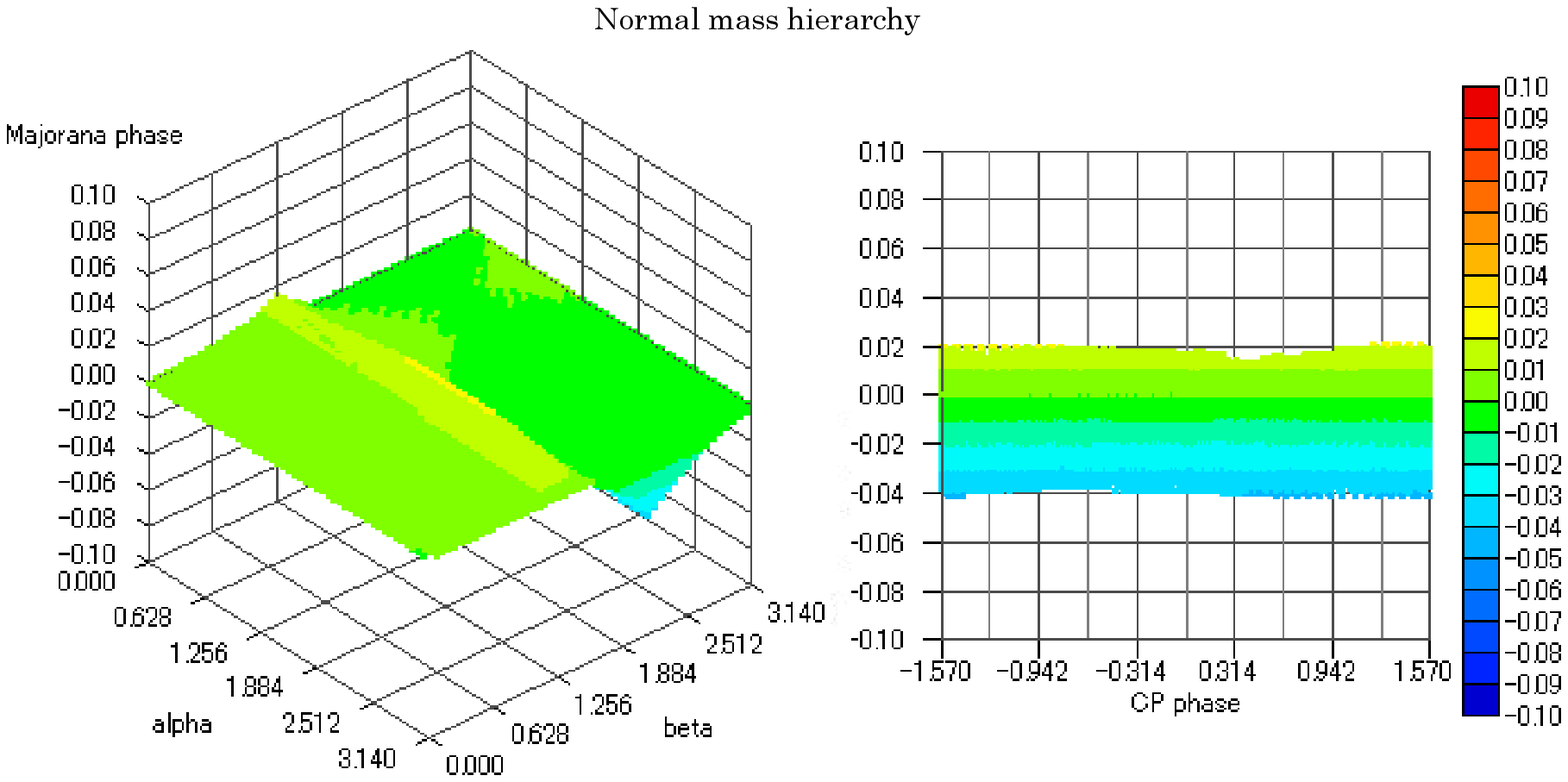}
\end{flushleft}
\vspace{-4mm}
\caption{The prediction of the Majorana phase as a function of the CP phase.}
\label{Fig:majorana-normal}
\end{figure}

\begin{figure}[!htbp]
\begin{flushleft}
\includegraphics*[20mm,162mm][195mm,263mm]{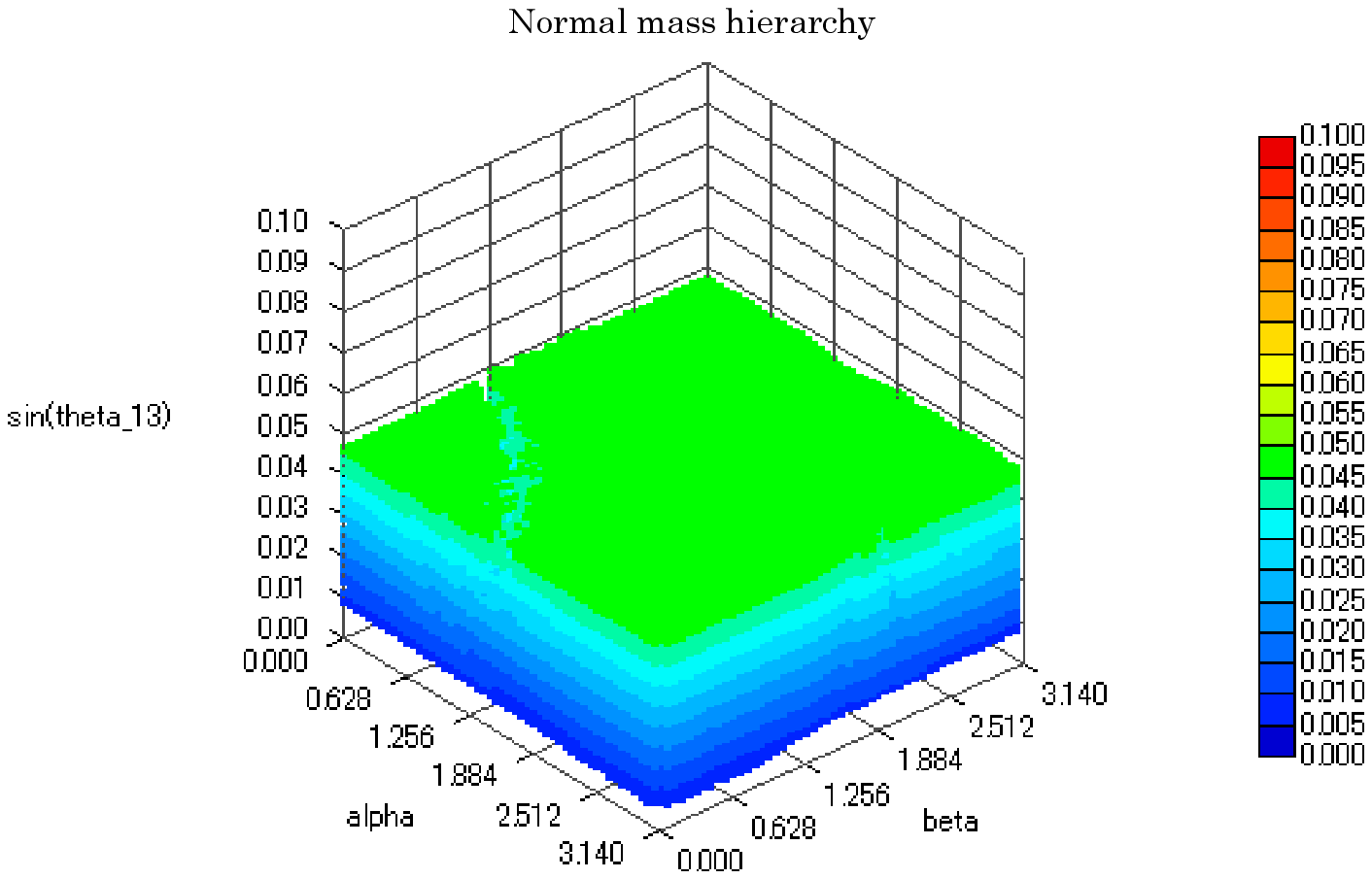}
\end{flushleft}
\vspace{-4mm}
\caption{The same as in FIG.\ref{Fig:phase-normal} but for $\sin\theta_{13}$.}
\label{Fig:sin13-normal}
\end{figure}

\begin{figure}[!htbp]
\begin{flushleft}
\includegraphics*[19.0mm,162mm][195mm,263mm]{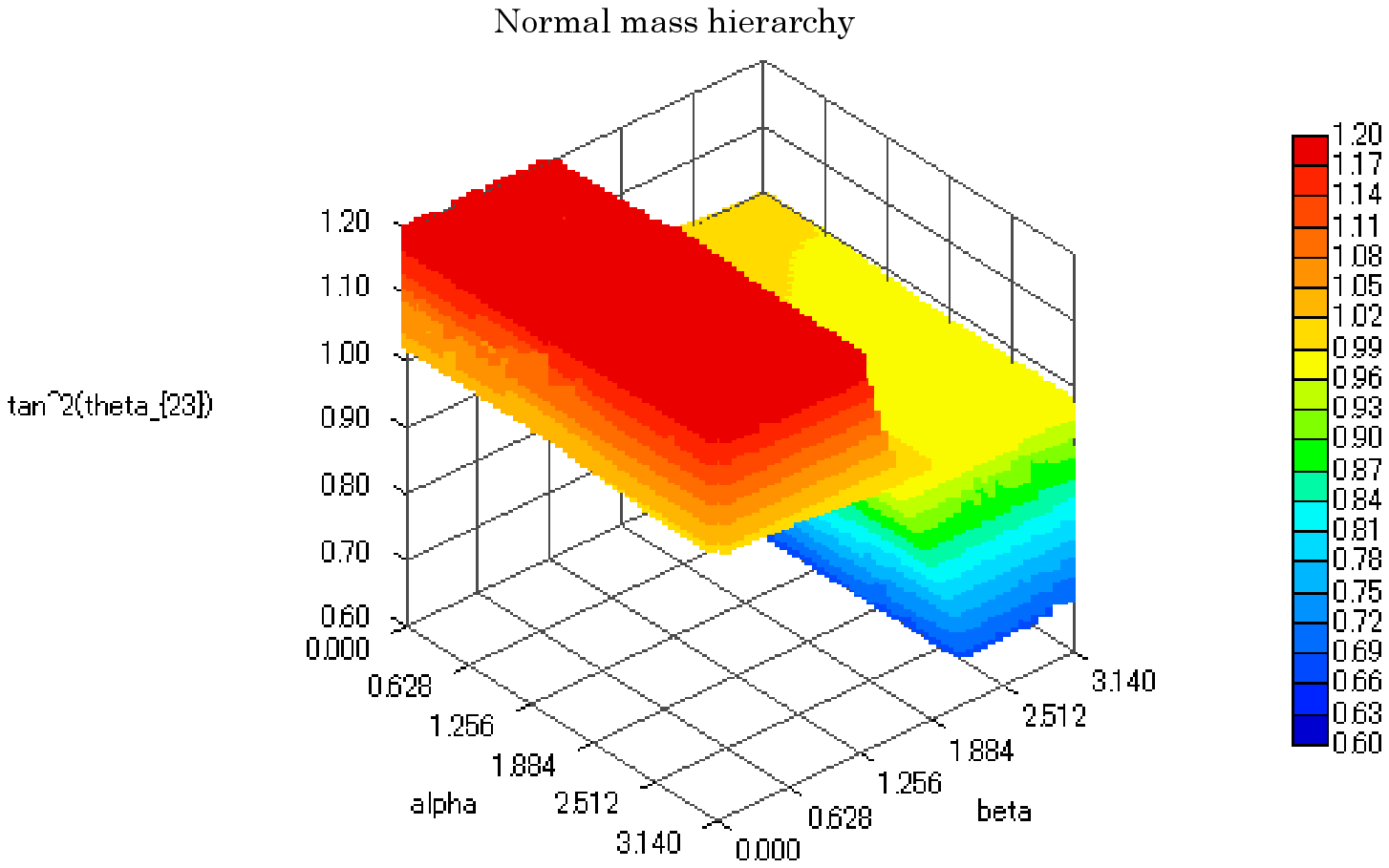}
\end{flushleft}
\vspace{-4mm}
\caption{The same as in FIG.\ref{Fig:phase-normal} but for $\tan^2\theta_{23}$.}
\label{Fig:tan23-normal}
\end{figure}

\begin{figure}[!htbp]
\begin{flushleft}
\includegraphics*[20mm,158mm][195mm,263mm]{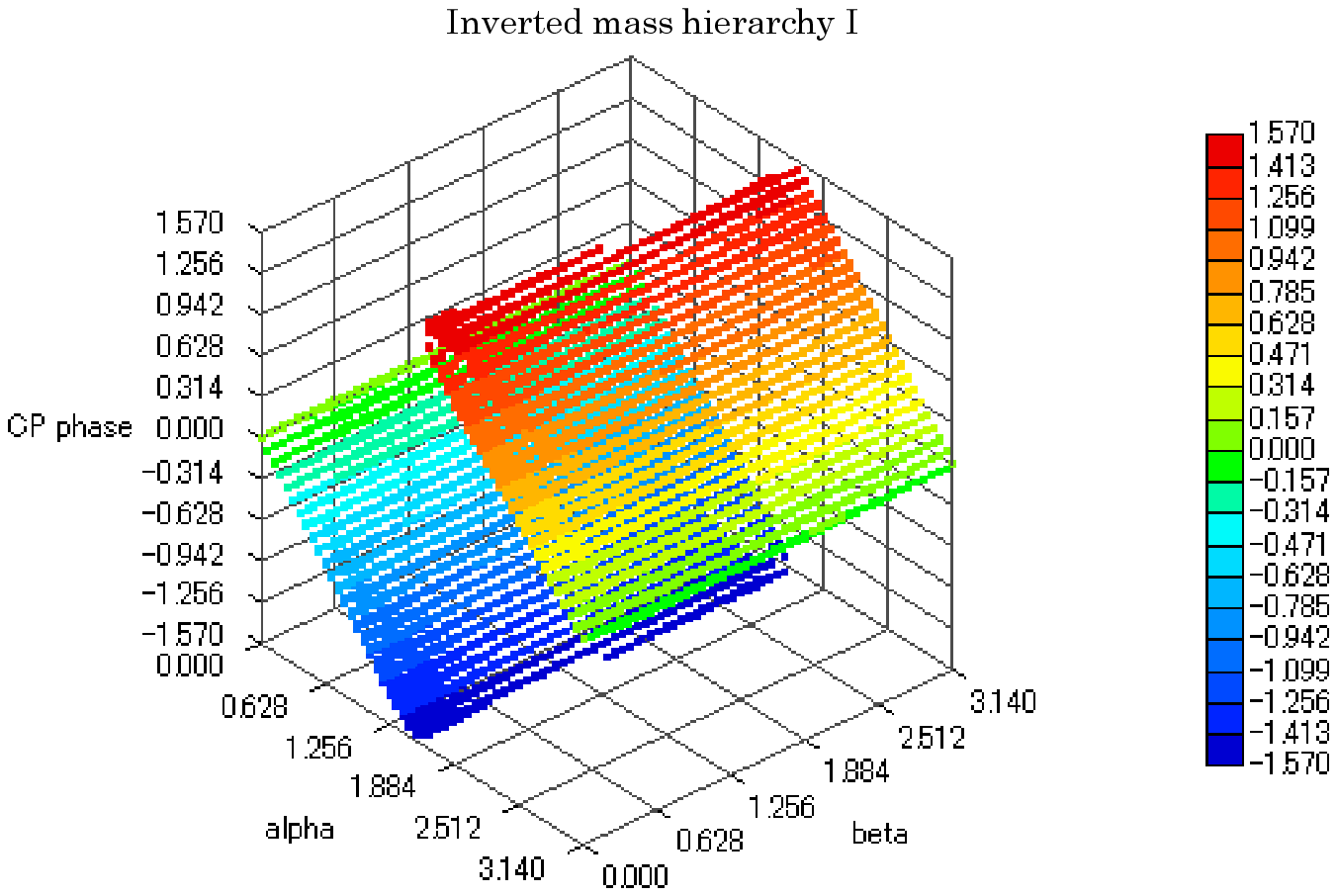}
\end{flushleft}
\vspace{-4mm}
\caption{The predictions of the Dirac CP phase $\delta+\rho$ as function of $\alpha$ and $\beta$ for the inverted mass hierarchy I.}
\label{Fig:phase-inverted1}
\end{figure}

\begin{figure}[!htbp]
\begin{flushleft}
\includegraphics*[13mm,161mm][205mm,263mm]{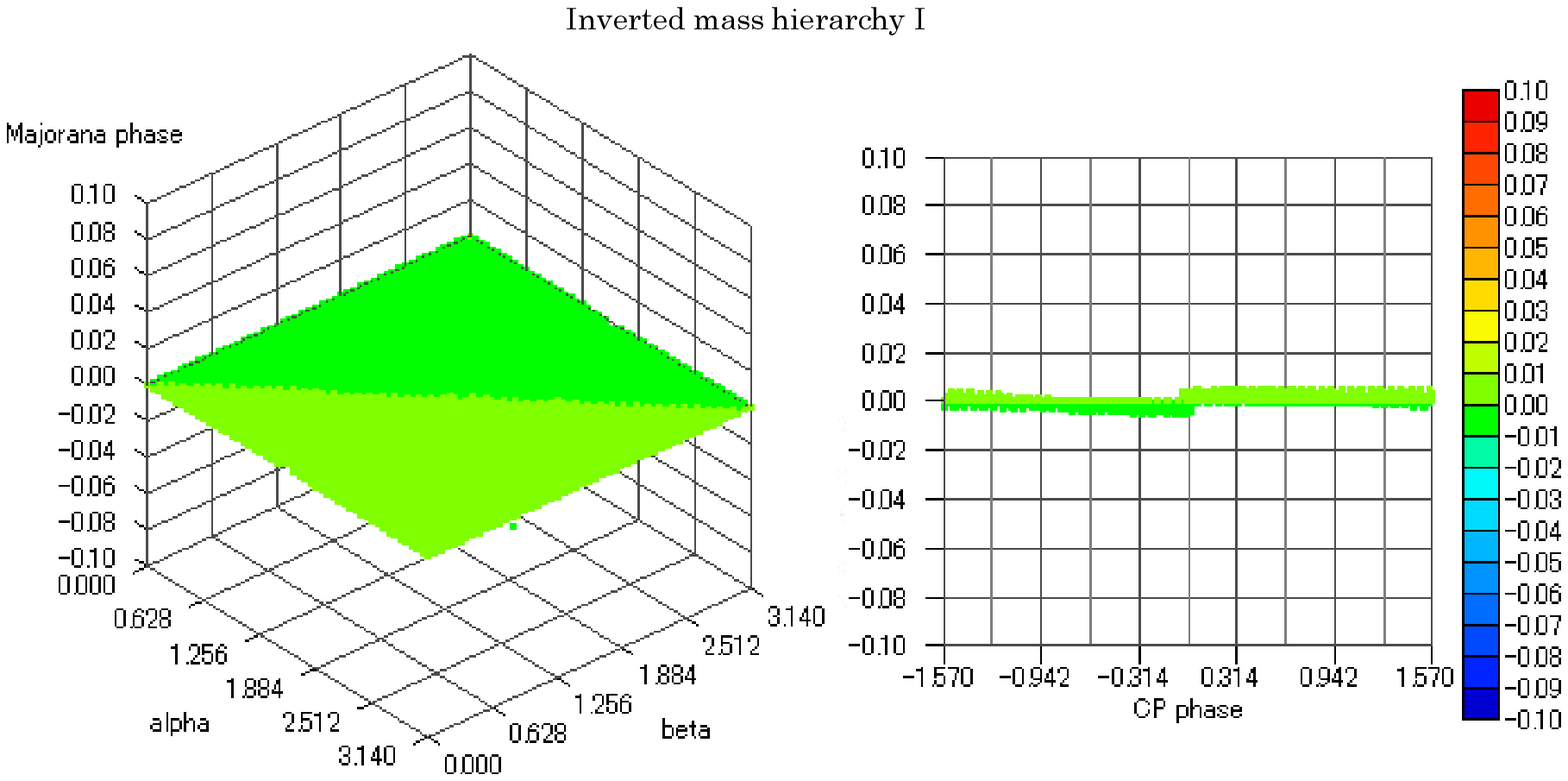}
\end{flushleft}
\vspace{-4mm}
\caption{The prediction of the Majorana phase as a function of the CP phase.}
\label{Fig:majorana-inverted1}
\end{figure}

\begin{figure}[!htbp]
\begin{flushleft}
\includegraphics*[20mm,162mm][195mm,263mm]{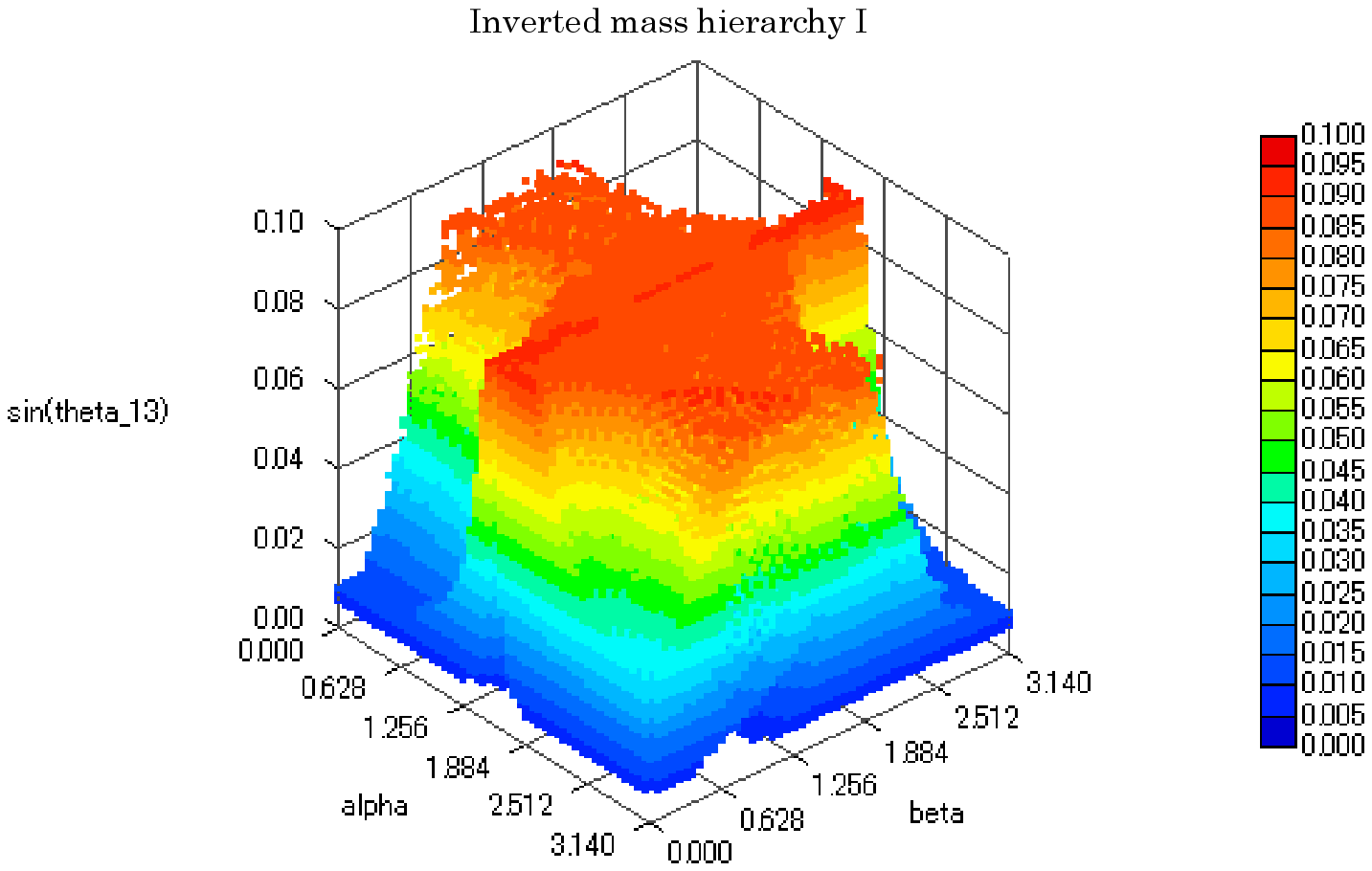}
\end{flushleft}
\vspace{-4mm}
\caption{The same as in FIG.\ref{Fig:phase-inverted1} but for $\sin\theta_{13}$.}
\label{Fig:sin13-inverted1}
\end{figure}

\begin{figure}[!htbp]
\begin{flushleft}
\includegraphics*[20mm,162mm][195mm,263mm]{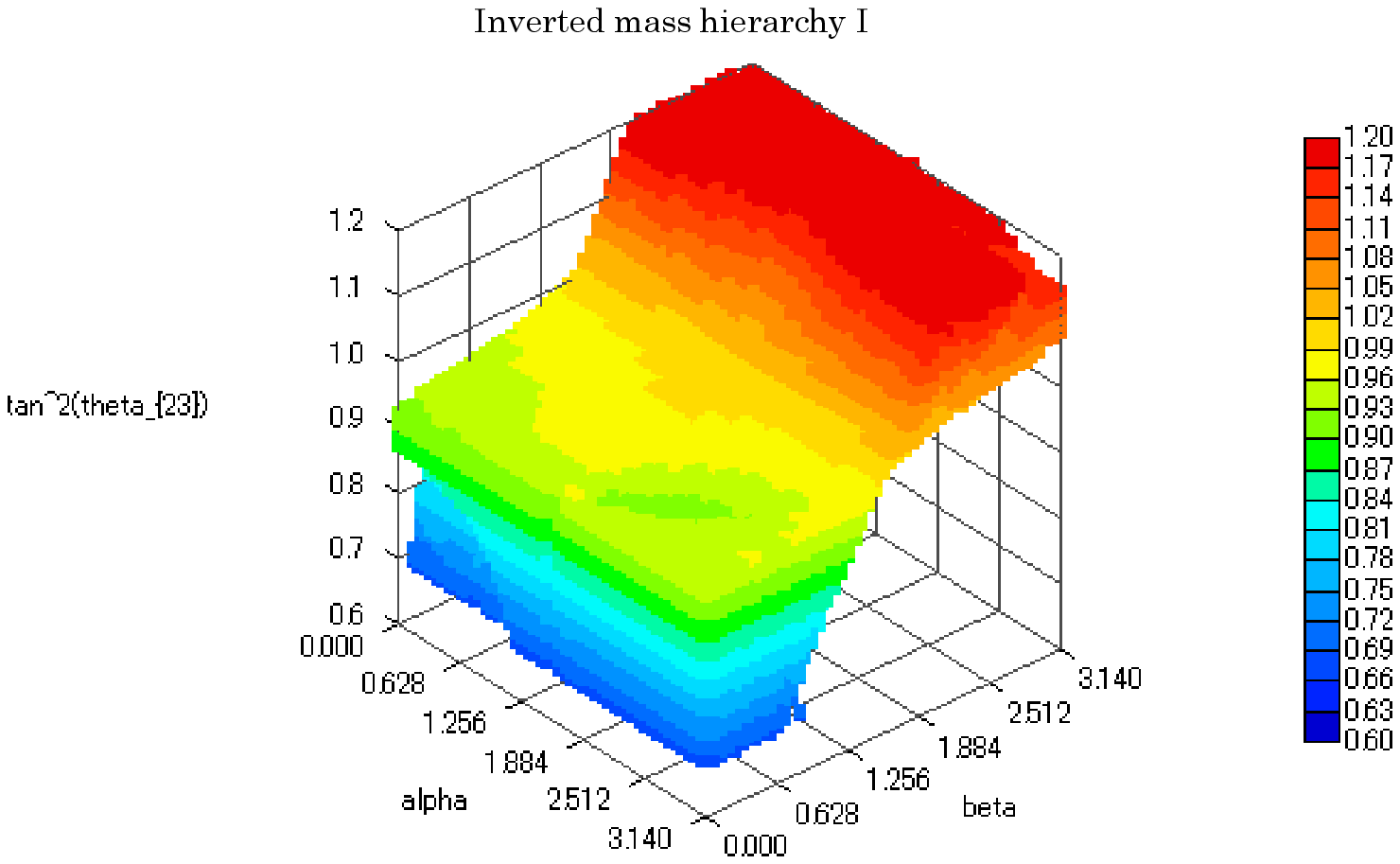}
\end{flushleft}
\vspace{-4mm}
\caption{The same as in FIG.\ref{Fig:phase-inverted1} but for $\tan^2\theta_{23}$.}
\label{Fig:tan23-inverted1}
\end{figure}

\begin{figure}[!htbp]
\begin{flushleft}
\includegraphics*[20mm,157mm][195mm,263mm]{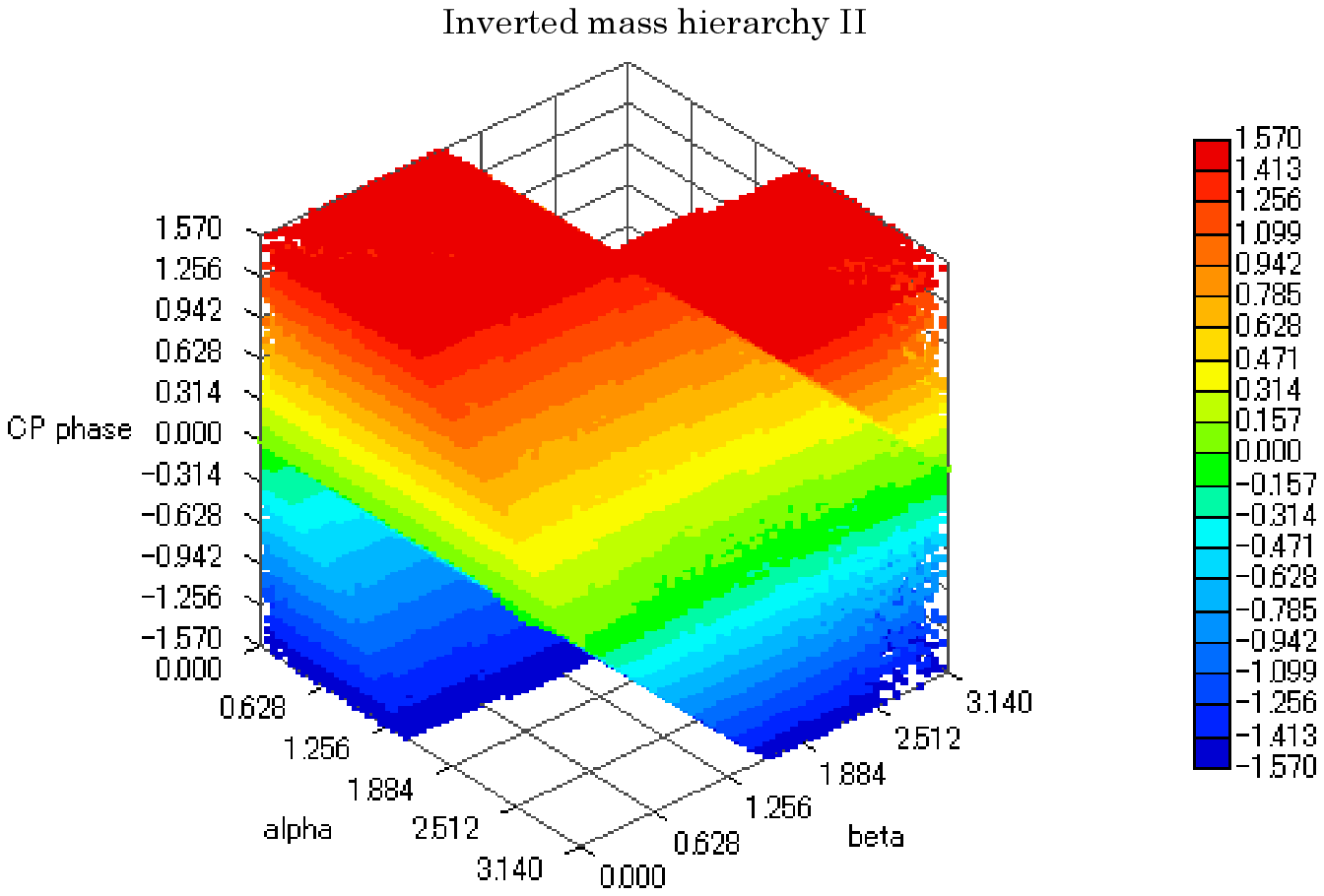}
\end{flushleft}
\vspace{-4mm}
\caption{The predictions of the Dirac CP phase $\delta+\rho$ as function of $\alpha$ and $\beta$ for the inverted mass hierarchy I\hspace{-.1em}I.}
\label{Fig:phase-inverted2}
\end{figure}

\begin{figure}[!htbp]
\begin{flushleft}
\includegraphics*[13mm,161mm][205mm,263mm]{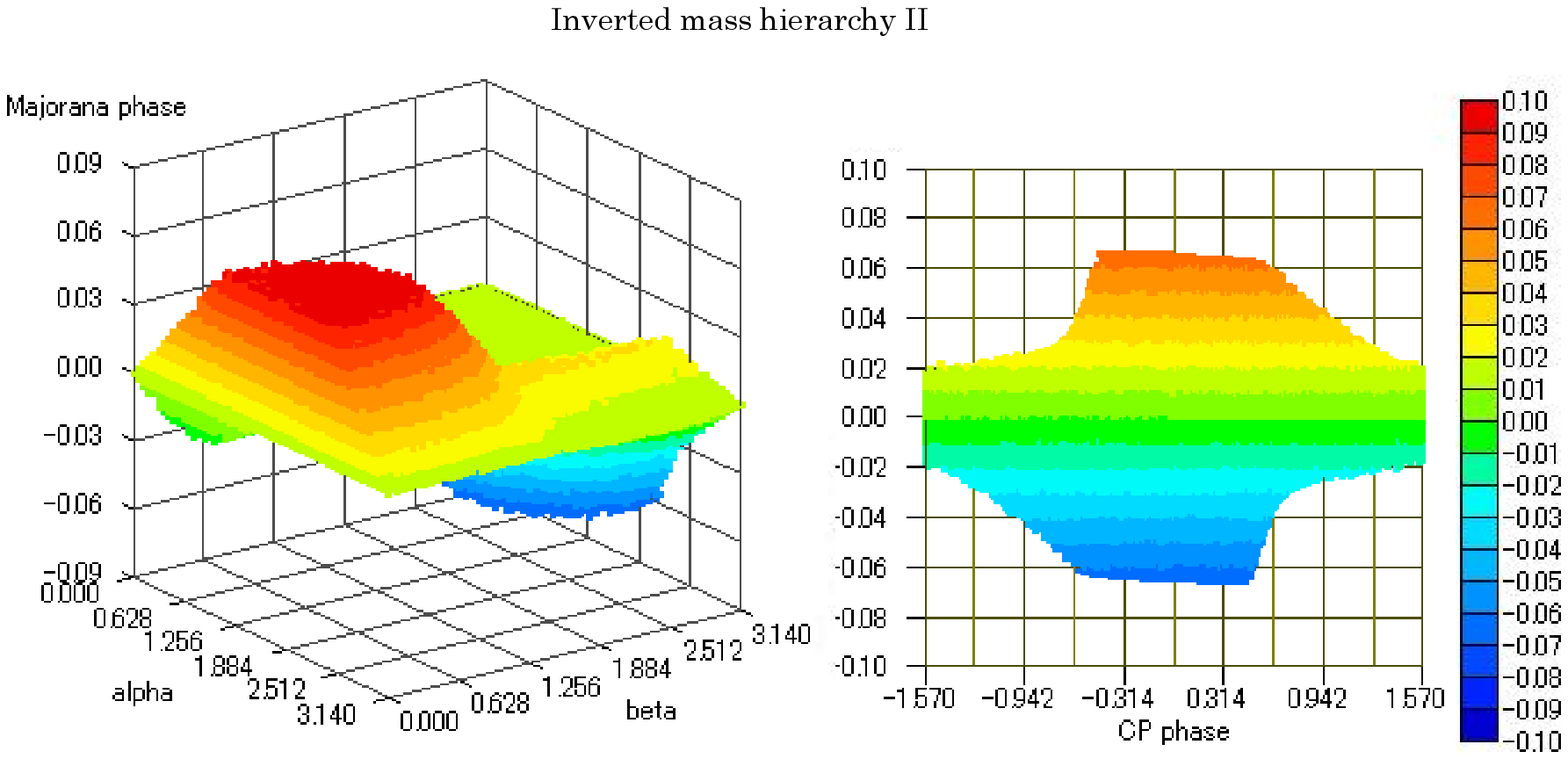}
\end{flushleft}
\vspace{-4mm}
\caption{The prediction of the Majorana phase as a function of the CP phase.}
\label{Fig:majorana-inverted2}
\end{figure}

\begin{figure}[!htbp]
\begin{flushleft}
\includegraphics*[20mm,161mm][200mm,263mm]{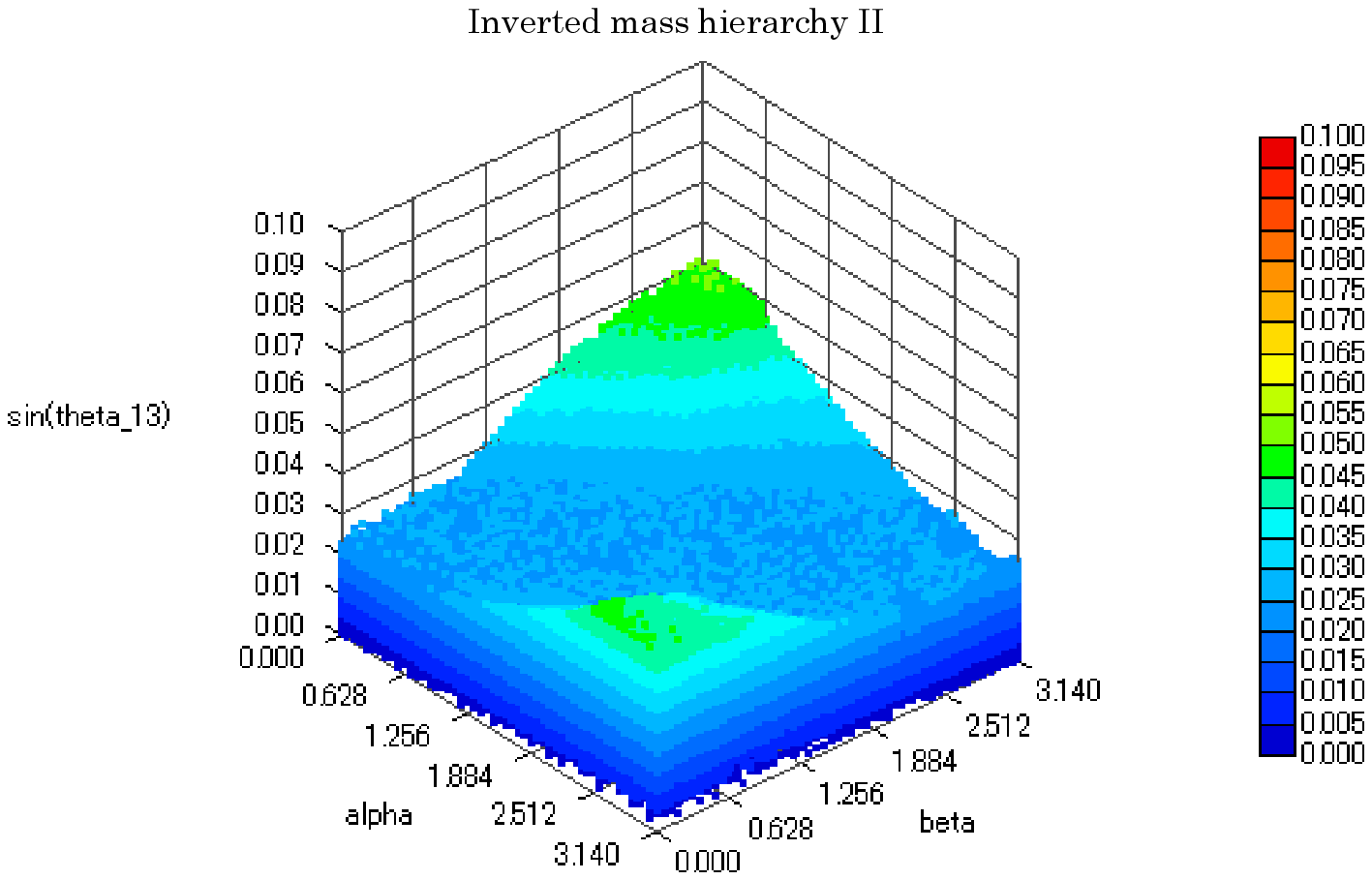}
\end{flushleft}
\vspace{-4mm}
\caption{The same as in FIG.\ref{Fig:phase-inverted2} but for $\sin\theta_{13}$.}
\label{Fig:sin13-inverted2}
\end{figure}

\begin{figure}[!htbp]
\begin{flushleft}
\includegraphics*[20mm,157mm][195mm,263mm]{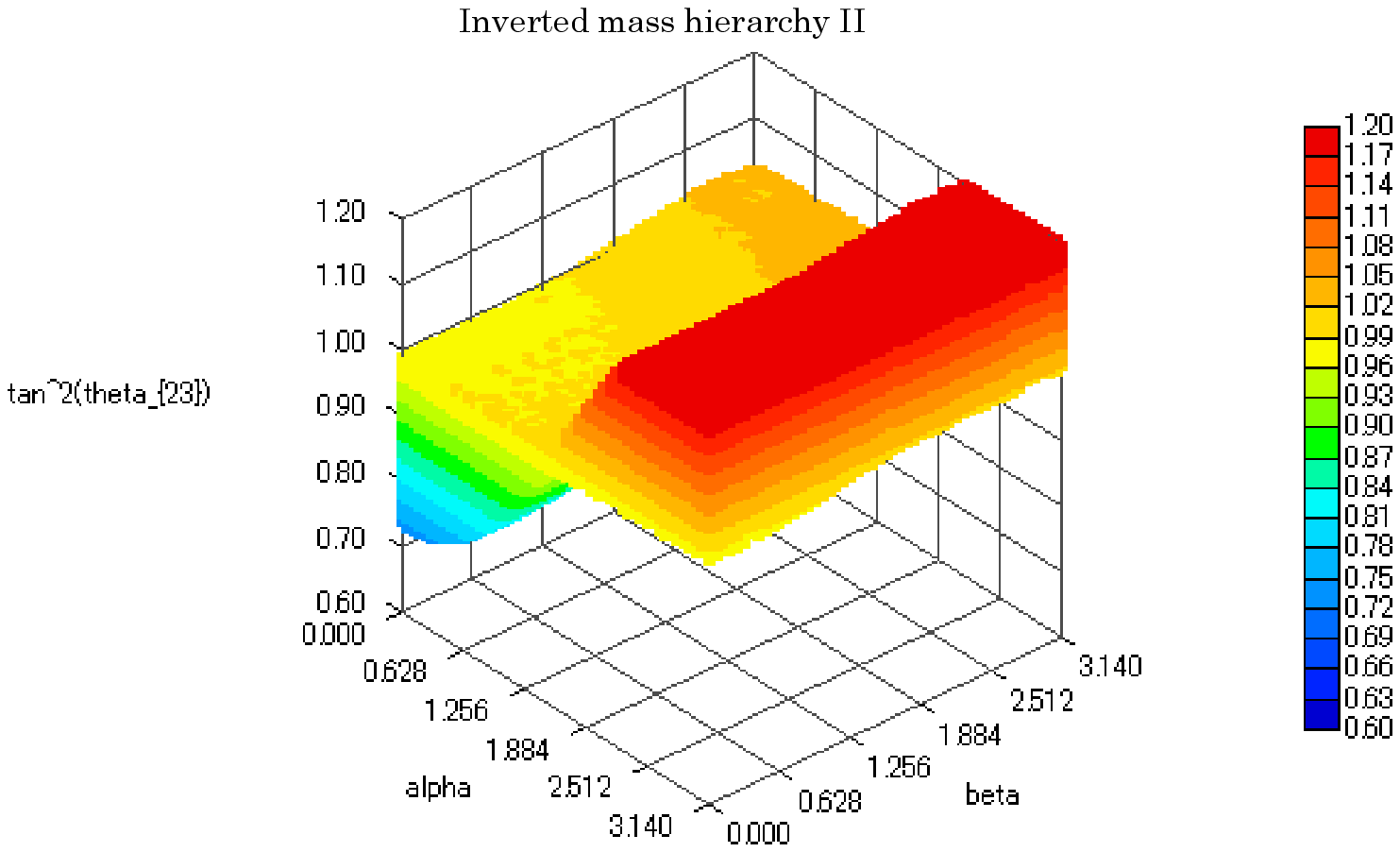}
\end{flushleft}
\vspace{-4mm}
\caption{The same as in FIG.\ref{Fig:phase-inverted2} but for $\tan^2\theta_{23}$.}
\label{Fig:tan23-inverted2}
\end{figure}

\begin{figure}[!htbp]
\begin{flushleft}
\includegraphics*[13mm,183mm][195mm,263mm]{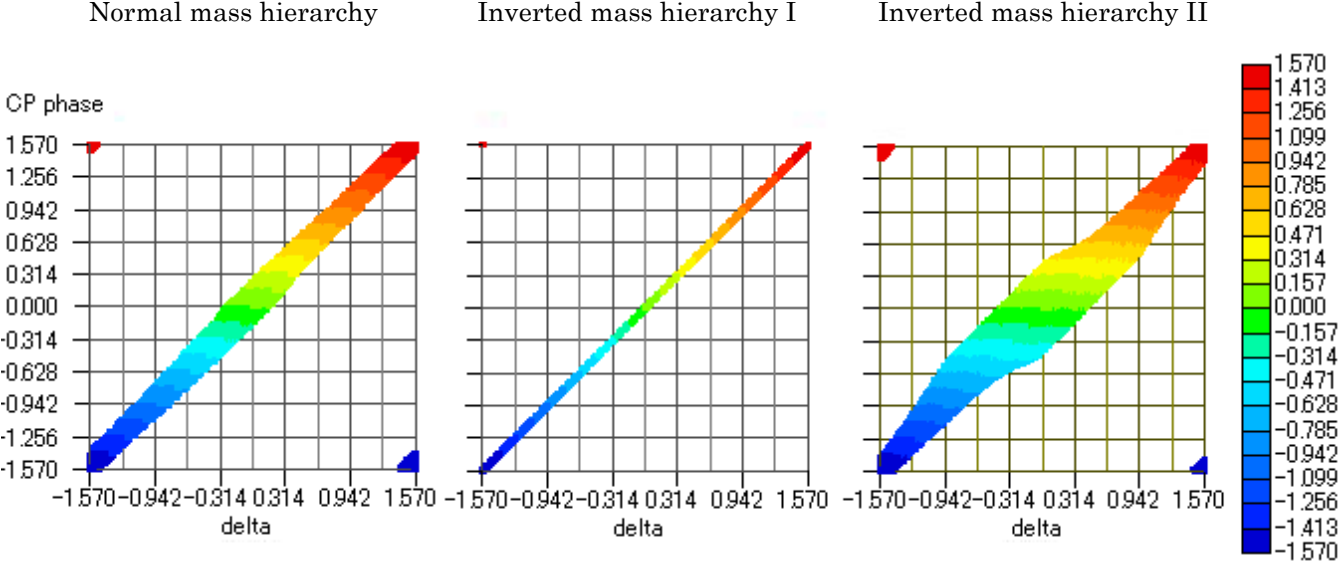}
\end{flushleft}
\vspace{-4mm}
\caption{The $\delta$-dependence of Dirac CP phase.}
\label{Fig:delta-phase}
\end{figure}

The predictions are depicted in FIG.\ref{Fig:phase-normal}-FIG.\ref{Fig:rho-majorana} for the 
normal and inverted mass hierarchies.  The effect of the sign of $\sigma$ is irrelevant because 
it always accompanies $\sin\theta_{13}$.  The gross features of the figures 
for the Dirac CP-violating phase $\delta_{CP}$
 accord with our results Eqs.(\ref{Eq:NormalPhases}), (\ref{Eq:Inverted1Phases}) and (\ref{Eq:Inverted2Phases}).
Namely, 
\begin{enumerate}
\item for the normal mass hierarchy, the crude proportionality of $\delta_{CP}$ to $\alpha$ shown in 
FIG.\ref{Fig:phase-normal} is accounted by Eq.(\ref{Eq:NormalPhases}) with $\rho\sim 0$ and 
the effect of $\rho$ gives scattered plots around the line $\delta_{CP}\propto\alpha$;
\item for the inverted mass hierarchy I (with $m_1\sim m_2$), the clear proportionality of 
$\delta_{CP}$ to $\alpha$ is shown in FIG.\ref{Fig:phase-inverted1} as suggested by 
Eq.(\ref{Eq:Inverted1Phases});
\item for the inverted mass hierarchy I\hspace{-.1em}I (with $m_1\sim -m_2$), the proportionality of 
$\delta_{CP}$ to $\beta$ can be seen as sharp edges in FIG.\ref{Fig:phase-inverted2} and 
is suggested by Eq.(\ref{Eq:Inverted1Phases}) for the region of $b^\prime_0\sim 0$;
\item In FIG.\ref{Fig:delta-phase}, the Dirac CP-violating phase is found to be proportional to $\delta$.  This 
behavior indicates that $\rho \sim 0$.  This is because 
$X$ in Eq.(\ref{Eq:Mdagger-ExactMixingAngles23}) starts with the $\mu$-$\tau$ symmetric contribution, 
which can be taken to be real, and, then, the phase $\rho$ starts with the $\mu$-$\tau$ breaking contribution, 
which generically suppressed, giving $\rho\sim 0$.
\end{enumerate}
The CP-violating Majorana phases $\phi$ are predicted
\begin{enumerate}
\item in FIG.\ref{Fig:majorana-normal}, FIG.\ref{Fig:majorana-inverted1} 
and FIG.\ref{Fig:majorana-inverted2}, where  
the CP-violating Majorana phase almost vanishes for the inverted mass hierarchy I as predicted 
in Eq.(\ref{Eq:Inverted1PredictionsMasses});
\item in FIG.\ref{Fig:rho-majorana}, where the CP-violating Majorana phase is found to be proportional to $\rho$.  
This feature can be roughly understood because of $\rho\sim 0$ in Eq.(\ref{Eq:M}) 
and the contribution of $\delta$ in the difference of Majorana phases almost vanish. Namely, we 
can estimate that $\phi \propto \rho$.  More precisely, our predictions Eqs.(\ref{Eq:Normal-phi}), 
(\ref{Eq:inverted-phi}) and (\ref{Eq:Inverted2-phi}) on $\phi$ are consistent 
with the behavior of these figures.
\end{enumerate}
The mixing angle $\theta_{13}$ satisfies the constraints:
\begin{enumerate}
\item $\sin\theta_{13}\mapleq 0.05$ for the normal mass hierarchy;
\item $\sin\theta_{13}\mapleq 0.09$ for the inverted mass hierarchy I (with $m_1\sim m_2$);
\item $\sin\theta_{13}\mapleq 0.05$ and $\sin\theta_{13}\sim 0.05$ around $\alpha\sim\beta\sim \pi$
for the inverted mass hierarchies I\hspace{-.1em}I (with $m_1\sim -m_2$),
\end{enumerate}
as can be seen from FIG.\ref{Fig:sin13-normal}, FIG.\ref{Fig:sin13-inverted1} 
and FIG.\ref{Fig:sin13-inverted2} and $\tan^2\theta_{23} > 1$
\begin{enumerate}
\item if $0\leq\beta\leq\pi/2$ for the normal mass hierarchy;
\item if $\pi/2\leq\beta\leq\pi$ for the inverted mass hierarchy I (with $m_1\sim m_2$);
\item $\pi/2\leq\alpha\leq\pi$ for the inverted mass hierarchies I\hspace{-.1em}I (with $m_1\sim -m_2$),
\end{enumerate}
as in FIG.\ref{Fig:tan23-normal}, FIG.\ref{Fig:tan23-inverted1} 
and FIG.\ref{Fig:tan23-inverted2}.

\section{\label{sec:5}Summary}
We have estimated CP-violating phases as well as mixing angles in the approximately $\mu-\tau$ symmetric minimal 
seesaw model.  
When heavy neutrino mass terms are real,
we have shown that CP-violating phases are determined by $\mu-\tau$ symmetry breaking phases 
in the neutrino Yukawa couplings as long as heavy neutrinos are transformed under the discrete $\mu-\tau$ 
symmetry group. 
As a result, phases in the flavor neutrino masses are expressed in terms of two phases 
$\alpha$ and $\beta$ as given by Eq.(\ref{Eq:NuMatrixSimple}).  On the other hand, such a property is not a 
general one if heavy neutrinos are not transformed.  
We have assumed the same phases $\alpha$ and $\beta$ to compare our predictions.
Furthermore, we have found that the normal mass hierarchy is permitted if 
heavy neutrinos are subject to the $\mu-\tau$ symmetry giving
a constraint of
$M_{e\mu}^{(+)2} \approx M_{ee}M_{\mu\mu}$, which is used to exclude the inverted mass hierarchy
and that the inverted mass hierarchy is permitted if the heavy neutrinos are blind to 
the $\mu-\tau$ symmetry giving a constraint of
$M_{\mu\tau} \approx -\sigma M_{\mu\mu}^{(+)}$, which is used to exclude the normal mass hierarchy. 
The restriction on the mass hierarchy is a general consequence 
of approximately $\mu-\tau$ symmetric minimal seesaw models
as long as no phases are present in heavy neutrinos.

\begin{figure}[!htbp]
\begin{flushleft}
\includegraphics*[13mm,187mm][195mm,263mm]{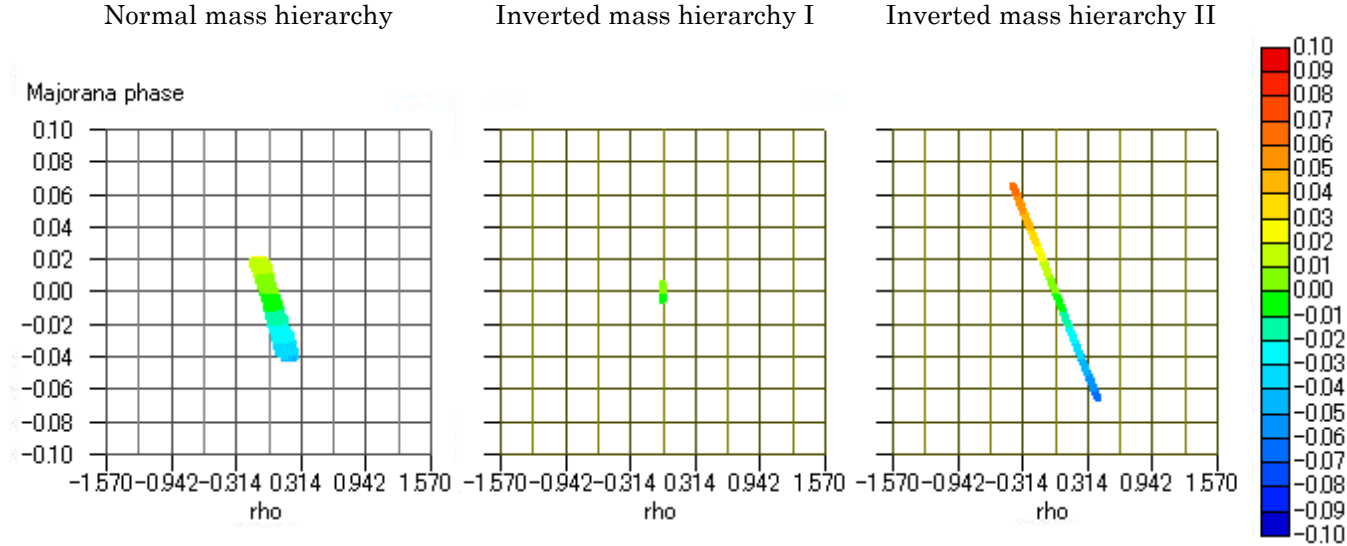}
\end{flushleft}
\vspace{-4mm}
\caption{The $\rho$-dependence of Majorana phase.}
\label{Fig:rho-majorana}
\end{figure}

We have also presented three textures, which give the consistent results with 
the current neutrino oscillation data: 
one describes the normal mass hierarchy as determined by Eq.(\ref{Eq:Normal-13-1}) 
and the other two describe the 
inverted mass hierarchy as determined by Eq.(\ref{Eq:Inverted-13-1}) and Eq.(\ref{Eq:Inverted-13-2}). 
Each textures have a small parameter $\eta$ to explain the smallness 
of the ratio of mass squared differences $\Delta m_\odot / \Delta m_{atm}(\equiv R)$, 
which is ${\mathcal O}(\sqrt{R})$ $({\mathcal O}(R))$ for the normal (inverted) mass hierarchy. 
The Dirac CP-violating phase is predicted from our formula Eq.(\ref{Eq:X-Y}) to yield $\delta_{CP} = \rho +\delta$.
Because of $\rho\sim 0$, we have found that 
the phase $\delta_{CP}$ is determined by $\alpha$ as $\delta_{CP} \approx \alpha$ 
as in Eq.(\ref{Eq:NormalPhases})
for the normal mass hierarchy and 
$\delta_{CP} \approx -\alpha$ for the inverted mass hierarchy I (with $m_1 \sim m_2$) 
as in Eq.(\ref{Eq:Inverted1Phases})
while 
$\delta_{CP}$ shows no dependence of $\alpha$ but a certain dependence of $\beta$ 
for the inverted mass hierarchy I\hspace{-.1em}I (with $m_1 \sim -m_2$) 
as in Eq.(\ref{Eq:Inverted2Phases}).
The numerical calculation is performed to make definite predictions, whose results are shown 
in FIG.\ref{Fig:phase-normal}-FIG.\ref{Fig:rho-majorana}. 
We have observed that
\begin{enumerate}
\item
The Dirac CP-phase $\delta_{CP}$ turns out to have a crude proportionality to $\alpha$ in the normal 
mass hierarchy as FIG.\ref{Fig:phase-normal}, a clear proportionality to 
$\alpha$ in the inverted mass hierarchy I (with 
$m_1 \sim m_2$) as FIG.\ref{Fig:phase-inverted1} and 
an proportionality to $\beta$ (for $b^\prime_0 \sim 0$) for inverted mass hierarchy I\hspace{-.1em}I (with 
$m_1 \sim -m_2$) as FIG.\ref{Fig:phase-inverted2}.
\item
The Majorana CP-violating phase $\phi$ is found to be suppressed since its main contributions 
arise from $\mu$-$\tau$ symmetry breaking terms and is estimated to be: 
$\phi\approx -\rho/4$ for the normal mass hierarchy, $\phi\approx 0$ 
for the inverted mass hierarchy I (with $m_1 \sim m_2$) and
$\phi\approx -\rho/6$ for the inverted mass hierarchy I\hspace{-.1em}I (with $m_1 \sim -m_2$) 
as in FIG.\ref{Fig:rho-majorana},
\item Our phases $\delta$ and $\rho$, respectively, yield main contributions 
to $\delta_{CP}$ and $\phi$ with $\rho\sim 0$ as in FIG.\ref{Fig:delta-phase} and \ref{Fig:rho-majorana}
, whose behaviors accord with our theoretical expectation.
\end{enumerate}
From these observations, we expect that the size of the CP-violating Majorana phase can be enhanced if we include 
the phase of the heavy neutrinos $\Theta_{+-}$ as in Eq.(\ref{Eq:HeavyMassMatrix}).  For 
the inverted mass hierarchy, we may relax our assumption that the $\mu$-$\tau$ symmetric terms are set 
to be real.  

Last but not least, we have to comment on the effective neutrino mass $m_{\beta\beta}$ \cite{TheoryMass-ee}
 used in the detection of the absolute neutrino mass \cite{AbsoluteMass}.  
 In our textures, $m_{\beta\beta}$ corresponds to the flavor mass of $e^{2i\rho}M_{ee}$  as in 
 Eq.(\ref{Eq:M-nu-ee-PDG}). 
 As stated in the Appendix, it is not $M_{ee}$ defined in Eq.(\ref{Eq:NewNuMatrix}) that can be 
 compared with experimental parameters, which are based on Eq.(\ref{Eq:Uu}).  
 In our case, since Eq.(\ref{Eq:Uu_general}) is an appropriate matrix, which should be transformed 
 into Eq.(\ref{Eq:Uu}).  In the course of this transformation, $M_{ee}$ in Eq.(\ref{Eq:NewNuMatrix}) is 
 changed to $e^{2i\rho}M_{ee}$, which is parameterized to be 
 $e^{2i\rho}a$ for a real $a$.  Therefore, in principle the phase $\rho$ has a chance 
 to be measured.  It is known that 
$\vert m_{\beta\beta}\vert $ is suppressed for the normal mass hierarchy, 
where the suppression 
factor $\eta$ appears in our texture, and is estimated to be 
$a(\sim\eta m_0)\sim\sqrt{\Delta m^2_\odot}$ with 
$\vert\Delta m^2_{atm}\vert \approx 4m^2_0$ 
while, for the inverted mass hierarchy,   
$\vert m_{\beta\beta}\vert \approx 2 m_0$ with $\vert\Delta m^2_{atm}\vert \approx m^2_0$
are obtained.

The predicted behaviors of CP phases are those at the seesaw scale. Radiative corrections to CP-phases should be 
evaluated to yield their observed values at the low-energy scale. Since these corrections are expected to be 
significant for the inverted mass hierarchy, we will estimate these corrections in the future publication. 
Furthermore since we know CP phases of the Yukawa couplings of neutrinos that can be inferred from the predicted 
Dirac and Majorana CP-violating phases, we can discuss how the leptogenesis is realized without referring to a 
specific from of flavor neutrino mass matrix but only with referring to more general framework of the $\mu-\tau$ 
symmetry breaking.

\vspace{3mm}
\noindent
\centerline{\small \bf ACKNOWLEGMENTS}

The authors are grateful to T. Kitabayashi for useful advices.

\appendix
\section{\label{sec:Appendix1}Flavor Neutrino Masses from Seesaw Mechanism}
In this Appendix, we evaluate  flavor neutrino masses $M_{ij}$ ($i,j=e,\mu,\tau$) 
that form a mass matrix $M_\nu$:
\begin{eqnarray}
&& M_\nu = \left( {\begin{array}{*{20}c}
	M_{ee} & M_{e\mu} & M_{e\tau}  \\
	M_{e\mu} & M_{\mu\mu} & M_{\mu\tau}  \\
	M_{e\tau} & M_{\mu\tau} & M_{\tau\tau}  \\
\end{array}} \right) = M^{(+)}+M^{(-)}.
\label{Eq:NewNuMatrix}
\end{eqnarray}
with
\begin{eqnarray}
&&
M^{(+)}  = \left( \begin{array}{*{20}c}
   M_{ee} & M^{(+)}_{e\mu } & - \sigma M^{(+)}_{e\mu }  \\
   M^{(+)}_{e\mu } & M^{(+)}_{\mu\mu } & M_{\mu \tau }   \\
    - \sigma M^{(+)}_{e\mu } & M_{\mu \tau } & M^{(+)}_{\mu\mu }\\
\end{array} \right),
\quad
M^{(-)}  = \left( \begin{array}{*{20}c}
   0 & M^{(-)}_{e\mu } & \sigma M^{(-)}_{e\mu }  \\
   M^{(-)}_{e\mu }& M^{(-)}_{\mu\mu } & 0  \\
   \sigma M^{(-)}_{e\mu } & 0 & - M^{(-)}_{\mu\mu } \\
\end{array} \right),
\label{Eq:Mnu-mutau-separation-2}
\end{eqnarray}
where $M^{(\pm)}_{e\mu} = (M_{e\mu} \mp \sigma M_{e\tau})/2$ and $M^{(\pm)}_{\mu\mu} = (M_{\mu\mu} \pm M_{\tau\tau})/2$.
This decomposition is just an identity.  However, it is so arranged that $M^{(+)}$ is invariant under
the interchange $\nu_\mu\leftrightarrow -\sigma\nu_\tau$.

After the Higgses develop vacuum expectation values, the seesaw mechanism gives
\begin{eqnarray}
&& 
 M_{ee}  =  - v^2 \left[ {\left( {ch_{ + e}  - se^{ - i\omega } h_{ - e} } \right)^2 \tilde M_1^{ - 1}  + \left( {se^{i\omega } h_{ + e}  + ch_{ - e} } \right)^2 \tilde M_2^{ - 1} } \right], 
\nonumber\\
&& 
 M_{e\mu }  =  - v^2 \left[ {\left( {ch_{ + e}  - se^{ - i\omega } h_{ - e} } \right)\left( {ch_{ + \mu }  - se^{ - i\omega } h_{ - \mu } } \right)\tilde M_1^{ - 1}  + \left( {se^{i\omega } h_{ + e}  + ch_{ - e} } \right)\left( {se^{i\omega } h_{ + \mu }  + ch_{ - \mu } } \right)\tilde M_2^{ - 1} } \right], 
\nonumber\\
&& 
 M_{e\tau }  =  - v^2 \left[ {\left( {ch_{ + e}  - se^{ - i\omega } h_{ - e} } \right)\left( {ch_{ + \tau }  - se^{ - i\omega } h_{ - \tau } } \right)\tilde M_1^{ - 1}  + \left( {se^{i\omega } h_{ + e}  + ch_{ - e} } \right)\left( {se^{i\omega } h_{ + \tau }  + ch_{ - \tau } } \right)\tilde M_2^{ - 1} } \right], 
\nonumber\\
&& 
 M_{\mu \mu }  =  - v^2 \left[ {\left( {ch_{ + \mu }  - se^{ - i\omega } h_{ - \mu } } \right)^2 \tilde M_1^{ - 1}  + \left( {se^{i\omega } h_{ + \mu }  + ch_{ - \mu } } \right)^2 \tilde M_2^{ - 1} } \right], 
\nonumber\\
&& 
 M_{\tau \tau }  =  - v^2 \left[ {\left( {ch_{ + \tau }  - se^{ - i\omega } h_{ - \tau } } \right)^2 \tilde M_1^{ - 1}  + \left( {se^{i\omega } h_{ + \tau }  + ch_{ - \tau } } \right)^2 \tilde M_2^{ - 1} } \right], 
\nonumber\\
&& 
 M_{\mu \tau }  =  - v^2 \left[ {\left( {ch_{ + \mu }  - se^{ - i\omega } h_{ - \mu } } \right)\left( {ch_{ + \tau }  - se^{ - i\omega } h_{ - \tau } } \right)\tilde M_1^{ - 1}  + \left( {se^{i\omega } h_{ + \mu }  + ch_{ - \mu } } \right)\left( {se^{i\omega } h_{ + \tau }  + ch_{ - \tau } } \right)\tilde M_2^{ - 1} } \right], 
\label{Eq:NuMatrixEntries}
\end{eqnarray}
where ${\tilde M}_ \pm^{-1} = M_\pm^{-1} e^{\pm i\varphi}$, $c=\cos\theta$, $s=\sin\theta$, and $v=\langle 0 \vert H_{u1}\vert 0\rangle$ for 
$H_u = (H_{u1}, H_{u2})^T$.
It is not difficult to demonstrate that Eq.(\ref{Eq:NuMatrixEntries}) satisfies $\det(M_\nu)=0$, 
which indicates the known property that the minimal seesaw model has one massless neutrino. 
The Yukawa couplings of $h_{\pm i}$ ($i=e,\mu,\tau$) literally represent the couplings to $N_\pm$. 
Therefore, the couplings of $h_{\pm i}$ should be expressed by the original Yukawa 
couplings defined in Eq.(\ref{Eq:SuperPotential}).
For example, in the case of $N=(N_\mu, N_\tau)$, 
we obtain that
\begin{eqnarray}
&& 
 h_{ + e}  = \frac{{h_{\mu e}  + \left( { - \sigma } \right)h_{\tau e} }}{{\sqrt 2 }},h_{ + \mu }  = \frac{{h_{\mu \mu }  + \left( { - \sigma } \right)h_{\tau \mu } }}{{\sqrt 2 }},{\rm{ }}h_{ + \tau }  = \frac{{h_{\mu \tau }  + \left( { - \sigma } \right)h_{\tau \tau } }}{{\sqrt 2 }},
\nonumber\\
&&
 h_{ - e}  = \frac{{h_{\mu e}  - \left( { - \sigma } \right)h_{\tau e} }}{{\sqrt 2 }},{\rm{ }}h_{ - \mu }  = \frac{{h_{\mu \mu }  - \left( { - \sigma } \right)h_{\tau \mu } }}{{\sqrt 2 }},{\rm{ }}h_{ - \tau }  = \frac{{h_{\mu \tau }  - \left( { - \sigma } \right)h_{\tau \tau } }}{{\sqrt 2 }},
\label{Eq:NuMatrixEntriesSpecific}
\end{eqnarray}
where the original Yukawa couplings are $h_{ij}$ ($i=\mu,\tau$, $j=e,\mu,\tau$).
 
\subsection{\label{sec:Appendix1-1}$N$ subject to the $\mu$-$\tau$ Symmetry}
In terms of Eq.(\ref{Eq:mu-tauMassMatrix-1}), Eq.(\ref{Eq:NuMatrixEntries}) is expressed as,
\begin{eqnarray}
&& 
 M_{ee}  =  - v^2 \left[ {\left( {ch_{ + e}^{\left(  +  \right)}  - se^{ - i\omega } h_{ - e}^{\left(  -  \right)} } \right)^2 \tilde M_1^{ - 1}  + \left( {se^{i\omega } h_{ + e}^{\left(  +  \right)}  + ch_{ - e}^{\left(  -  \right)} } \right)^2 \tilde M_2^{ - 1} } \right],
\nonumber\\
&& 
 M_{e\mu }^{\left(  +  \right)}  =  - v^2 \left[ {\left( {ch_{ + e}^{\left(  +  \right)}  - se^{ - i\omega } h_{ - e}^{\left(  -  \right)} } \right)\left( {ch_{ + \mu }^{\left(  +  \right)}  - se^{ - i\omega } h_{ - \mu }^{\left(  -  \right)} } \right)\tilde M_1^{ - 1}  + \left( {se^{i\omega } h_{ + e}^{\left(  +  \right)}  + ch_{ - e}^{\left(  -  \right)} } \right)\left( {se^{i\omega } h_{ + \mu }^{\left(  +  \right)}  + ch_{ - \mu }^{\left(  -  \right)} } \right)\tilde M_2^{ - 1} } \right],
\nonumber\\
&& 
 M_{e\mu }^{\left(  -  \right)}  =  - v^2 \left[ {\left( {ch_{ + e}^{\left(  +  \right)}  - se^{ - i\omega } h_{ - e}^{\left(  -  \right)} } \right)\left( {ch_{ + \mu }^{\left(  -  \right)}  - se^{ - i\omega } h_{ - \mu }^{\left(  +  \right)} } \right)\tilde M_1^{ - 1}  + \left( {se^{i\omega } h_{ + e}^{\left(  +  \right)}  + ch_{ - e}^{\left(  -  \right)} } \right)\left( {se^{i\omega } h_{ + \mu }^{\left(  -  \right)}  + ch_{ - \mu }^{\left(  +  \right)} } \right)\tilde M_2^{ - 1} } \right],
\nonumber\\
&& 
 M_{\mu \mu }^{\left(  +  \right)}  =  - v^2 \left[ \begin{array}{l}
 \left( {\left( {ch_{ + \mu }^{\left(  +  \right)}  - se^{ - i\omega } h_{ - \mu }^{\left(  -  \right)} } \right)^2  + \left( {ch_{ + \mu }^{\left(  -  \right)}  - se^{ - i\omega } h_{ - \mu }^{\left(  +  \right)} } \right)^2 } \right)\tilde M_1^{ - 1}  \\ 
  + \left( {\left( {ch_{ - \mu }^{\left(  -  \right)}  + se^{i\omega } h_{ + \mu }^{\left(  +  \right)} } \right)^2  + \left( {ch_{ - \mu }^{\left(  +  \right)}  + se^{i\omega } h_{ + \mu }^{\left(  -  \right)} } \right)^2 } \right)\tilde M_2^{ - 1}  \\ 
 \end{array} \right],
\nonumber\\
&& 
 M_{\mu \mu }^{\left(  -  \right)} {\rm{ }} =  - 2v^2 \left[ {\left( {ch_{ + \mu }^{\left(  +  \right)}  - se^{ - i\omega } h_{ - \mu }^{\left(  -  \right)} } \right)\left( {ch_{ + \mu }^{\left(  -  \right)}  - se^{ - i\omega } h_{ - \mu }^{\left(  +  \right)} } \right)\tilde M_1^{ - 1}  + \left( {ch_{ - \mu }^{\left(  -  \right)}  + se^{i\omega } h_{ + \mu }^{\left(  +  \right)} } \right)\left( {ch_{ - \mu }^{\left(  +  \right)}  + se^{i\omega } h_{ + \mu }^{\left(  -  \right)} } \right)\tilde M_2^{ - 1} } \right],
\nonumber
\end{eqnarray}
\begin{eqnarray}
&& 
 M_{\mu \tau }  =  - v^2 \left( { - \sigma } \right)\left[ \begin{array}{l}
 \left( {\left( {ch_{ + \mu }^{\left(  +  \right)}  - se^{ - i\omega } h_{ - \mu }^{\left(  -  \right)} } \right)^2  - \left( {ch_{ + \mu }^{\left(  -  \right)}  - se^{ - i\omega } h_{ - \mu }^{\left(  +  \right)} } \right)^2 } \right)\tilde M_1^{ - 1}  \\ 
  + \left( {\left( {se^{i\omega } h_{ + \mu }^{\left(  +  \right)}  + ch_{ - \mu }^{\left(  -  \right)} } \right)^2  - \left( {ch_{ - \mu }^{\left(  +  \right)}  + se^{i\omega } h_{ + \mu }^{\left(  -  \right)} } \right)^2 } \right)\tilde M_2^{ - 1}  \\ 
 \end{array} \right],
\label{Eq:NuMatrixEntries-1}
\end{eqnarray}
where $c=\cos\theta$ and $s = \sin\theta$.  
The suffices $\pm$ represent for $(N_+, N_-)$ that have $N_\pm\rightarrow \pm N_\pm$ under the $\mu$-$\tau$ symmetry 
transformation.

\subsubsection{\label{appendix:1-1-1}$\mu$-$\tau$ Symmetry Breaking Case}
The approximate $\mu$-$\tau$ symmetry calls for 
\begin{eqnarray}
&& 
h_{- e}^{\left( - \right)}\approx 0, 
\quad
h_{+\mu}^{\left( - \right)}\approx 0,
\quad
h_{-\mu}^{\left( - \right)}\approx 0,
\label{Eq:FirstOrder1-1}
\end{eqnarray}
as well as ${\bf M}_{+- }\approx 0$, 
which yields
\begin{eqnarray}
&&
\cos\theta \approx 1,
\quad
\sin\theta \approx 
\frac{{{\bf M}_{R +  - } r}}{{{\bf M}_{R -  - }  - {\bf M}_{R +  + } }},
\label{Eq:FirstOrder1-2}
\end{eqnarray}
where
$r$ is defined in Eq.(\ref{Eq:HeavyMixingAngle-r}).
Using these approximations, we obtain Eq.(\ref{Eq:NuMatrixEntries-1}) up to the first order in the 
parameters of Eqs.(\ref{Eq:FirstOrder1-1}) and (\ref{Eq:FirstOrder1-2}): 
\begin{eqnarray}
&&
 M_{ee}  \approx  - v^2 h_{ + e}^{\left(  +  \right)2} \tilde M_1^{ - 1},
\nonumber\\
&&
 M_{e\mu }^{\left(  +  \right)}  \approx  - v^2 h_{ + e}^{\left(  +  \right)} h_{ + \mu }^{\left(  +  \right)} \tilde M_1^{ - 1},
\nonumber\\
&&
 M_{e\mu }^{\left(  -  \right)}  \approx  - v^2 \left[ {\left( {h_{ + \mu }^{\left(  -  \right)}  - se^{ - i\omega } h_{ - \mu }^{\left(  +  \right)} } \right)h_{ + e}^{\left(  +  \right)} \tilde M_1^{ - 1}  + \left( {h_{ - e}^{\left(  -  \right)}  + se^{i\omega } h_{ + e}^{\left(  +  \right)} } \right)h_{ - \mu }^{\left(  +  \right)} \tilde M_2^{ - 1} } \right], 
\nonumber\\
&&
 M_{\mu \mu }^{\left(  +  \right)}  \approx  - v^2 \left( {h_{ + \mu }^{\left(  +  \right)2} \tilde M_1^{ - 1}  + h_{ - \mu }^{\left(  +  \right)2} \tilde M_2^{ - 1} } \right),
\nonumber\\
&&
 M_{\mu \mu }^{\left(  -  \right)}  \approx  - 2v^2 \left[ {\left( {h_{ + \mu }^{\left(  -  \right)}  - se^{ - i\omega } h_{ - \mu }^{\left(  +  \right)} } \right)h_{ + \mu }^{\left(  +  \right)} \tilde M_1^{ - 1}  + \left( {h_{ - \mu }^{\left(  -  \right)}  + se^{i\omega } h_{ + \mu }^{\left(  +  \right)} } \right)h_{ - \mu }^{\left(  +  \right)} \tilde M_2^{ - 1} } \right],
\nonumber\\
&&
 M_{\mu \tau }  \approx  - v^2 \left( { - \sigma } \right)\left( {h_{ + \mu }^{\left(  +  \right)2} \tilde M_1^{ - 1}  - h_{ - \mu }^{\left(  +  \right)2} \tilde M_2^{ - 1} } \right),
\label{Eq:MassFlavorAppendix-1}
\end{eqnarray}
where $s \approx 0$.

\subsubsection{\label{appendix:1-1-2}$\mu$-$\tau$ Symmetric Case}
The $\mu$-$\tau$ symmetric textures containing one massless neutrino 
should describe either the normal mass hierarchy or the inverted mass hierarchy. 
 Imposing the conditions:
\begin{eqnarray}
&&
\cos\theta =1,
\quad
 \sin\theta = 0,
\label{Eq:HeavyMixing}
\end{eqnarray}
we obtain from Eq.(\ref{Eq:NuMatrixEntries-1}) that 
\begin{eqnarray}
&&
 M_{ee} \left( { = a_0 } \right) =  - v^2 h_{ + e}^{\left(  +  \right)2} \tilde M_1^{ - 1},
\nonumber\\
&&
 M_{e\mu } \left( { = b_0 } \right) =  - v^2 h_{ + e}^{\left(  +  \right)} h_{ + \mu }^{\left(  +  \right)} \tilde M_1^{ - 1},
\nonumber\\
&&
 M_{e\tau } \left( { = c_0 } \right) =  - \sigma M_{e\mu },
\nonumber\\
&&
 M_{\mu \mu } \left( { = d_0  \equiv d_ +   + d_ -  } \right) =  - v^2 \left( {h_{ + \mu }^{\left(  +  \right)2} \tilde M_1^{ - 1}  + h_{ - \mu }^{\left(  +  \right)2} \tilde M_2^{ - 1} } \right),
\nonumber\\
&&
 M_{\mu \tau } \left( { = e_0  \equiv  - \sigma \left( {d_ +   - d_ -  } \right)} \right) =  - v^2 \left( { - \sigma } \right)\left( {h_{ + \mu }^{\left(  +  \right)2} \tilde M_1^{ - 1}  - h_{ - \mu }^{\left(  +  \right)2} \tilde M_2^{ - 1} } \right),
\nonumber\\
&&
 M_{\tau \tau } \left( { = f_0 } \right) = M_{\mu \mu },
\label{Eq:NuMatrixEntriesSym-1}
\end{eqnarray}
This texture turns out to give  mass terms
\begin{eqnarray}
&&
\frac{1}{2}a_0 \nu _e \nu _e  + b_0 \frac{{\nu _e \nu _ +   + \nu _ +  \nu _e }}{{\sqrt 2 }} + d_ +  \nu _ +  \nu _ +   + d_ -  \nu _ -  \nu _ -  .
\label{Eq:NuMassTermSym}
\end{eqnarray}
This form of Eq.(\ref{Eq:NuMassTermSym}) is also valid for the model with $(N_e, N_-)$.  We then obtain
\begin{eqnarray}
&&
M_\nu = \left( {\begin{array}{*{20}c}
   {a_0 } & {b_0 } & { - \sigma b_0 }  \\
   {b_0 } & {d_ +   + d_ -  } & { - \sigma \left( {d_ +   - d_ -  } \right)}  \\
   { - \sigma b_0 } & { - \sigma \left( {d_ +   - d_ -  } \right)} & {d_ +   + d_ -  }  \\
\end{array}} \right),
\label{Eq:NuTextureSym-1}
\end{eqnarray}
where
\begin{eqnarray}
&&
a_0  =  - v^2 h_{ + e}^{\left(  +  \right)2} \tilde M_1^{ - 1} ,
\quad
b_0  =  - v^2 h_{ + e}^{\left(  +  \right)} h_{ + \mu }^{\left(  +  \right)} \tilde M_1^{ - 1},
\nonumber\\
&&
d_ +   =  - v^2 h_{ + \mu }^{\left(  +  \right)2} \tilde M_1^{ - 1} ,
\quad
d_ -   =  - v^2 h_{ - \mu }^{\left(  +  \right)2} \tilde M_2^{ - 1}.
\label{Eq:NuTextureParam-1}
\end{eqnarray}
from which we observe that
\begin{eqnarray}
&&
 b^2_0 = a_0d_+
\label{Eq:NuTextureParamConstraint}
\end{eqnarray}
is satisfied.  

To see how the 
mass hierarchies are realized, it is sufficient to check the ideal case, where $m_1=m_2=0$ with 
$m_3 \neq 0$ for the normal mass hierarchy and $m_1 = \pm m_2$ with $m_3 = 0$ for the inverted mass hierarchy.
The gross structure of $M_\nu$ for the inverted mass hierarchy is described by ideal textures:
\begin{eqnarray}
&& 
M^{(1)}_\nu=m_0\left( 
\begin{array}{*{20}c}
   2 & 0 & 0\\
   0 & 1 & -\sigma  \\
   0& -\sigma  & 1\\
\end{array}
\right),
\quad
M^{(2)}_\nu=m_0\left( 
\begin{array}{*{20}c}
   -2 & b_0 & -\sigma b_0\\
   b_0 & 1 & -\sigma  \\
   -\sigma b_0& -\sigma  & 1\\
\end{array}
\right)(b_0\neq 0),
\label{Eq:NuMatrixIdeal}
\end{eqnarray}
respectively, corresponding to $m_1= m_2 $ and $m_1=- m_2 $, which can be seen from Eq.(\ref{Eq:PredictionsMassSym}).
Since $d_+ = m_0$ and $d_- = 0$ should be satisfied,
we find that Eq.(\ref{Eq:NuTextureParamConstraint}) 
gives $0 = 2 m^2_0$ for $M^{(1)}_\nu$ and $b^2_0 = - 2 m^2_0$ for $M^{(2)}_\nu$. 
Therefore, the $M^{(1)}_\nu$ case is obviously ruled out and the $M^{(2)}_\nu$ case is allowed 
if $b_0$ is nearly pure imaginary.  Since no phases is present in the $N$ mass terms, $b_0$ is (almost) 
real and the $M^{(2)}_\nu$ case is also excluded. 

\subsection{\label{sec:Appendix1-2}$N$ blind to the $\mu$-$\tau$ Symmetry}
In terms of Eq.(\ref{Eq:mu-tauMassMatrix-2}), Eq.(\ref{Eq:NuMatrixEntries}) is expressed as
\begin{eqnarray}
&&
 M_{ee}  =  - v^2 \left[ {\left( {ch_{ + e}^{\left(  +  \right)}  - se^{ - i\omega } h_{ - e}^{\left(  +  \right)} } \right)^2 \tilde M_1^{ - 1}  + \left( {se^{i\omega } h_{ + e}^{\left(  +  \right)}  + ch_{ - e}^{\left(  +  \right)} } \right)^2 \tilde M_2^{ - 1} } \right], 
\nonumber\\
&&
 M_{e\mu }^{\left(  +  \right)}  =  - v^2 \left[ {\left( {ch_{ + e}^{\left(  +  \right)}  - se^{ - i\omega } h_{ - e}^{\left(  +  \right)} } \right)\left( {ch_{ + \mu }^{\left(  +  \right)}  - se^{ - i\omega } h_{ - \mu }^{\left(  +  \right)} } \right)\tilde M_1^{ - 1}  + \left( {se^{i\omega } h_{ + e}^{\left(  +  \right)}  + ch_{ - e}^{\left(  +  \right)} } \right)\left( {se^{i\omega } h_{ + \mu }^{\left(  +  \right)}  + ch_{ - \mu }^{\left(  +  \right)} } \right)\tilde M_2^{ - 1} } \right],
\nonumber\\
&&
 M_{e\mu }^{\left(  -  \right)}  =  - v^2 \left[ {\left( {ch_{ + e}^{\left(  +  \right)}  - se^{ - i\omega } h_{ - e}^{\left(  +  \right)} } \right)\left( {ch_{ + \mu }^{\left(  -  \right)}  - se^{ - i\omega } h_{ - \mu }^{\left(  -  \right)} } \right)\tilde M_1^{ - 1}  + \left( {se^{i\omega } h_{ + e}^{\left(  +  \right)}  + ch_{ - e}^{\left(  +  \right)} } \right)\left( {se^{i\omega } h_{ + \mu }^{\left(  -  \right)}  + ch_{ - \mu }^{\left(  -  \right)} } \right)\tilde M_2^{ - 1} } \right],
\nonumber\\
&&
 M_{\mu \mu }^{\left(  +  \right)}  =  - v^2 \left[ \begin{array}{l}
 \left( {\left( {ch_{ + \mu }^{\left(  +  \right)}  - se^{ - i\omega } h_{ - \mu }^{\left(  +  \right)} } \right)^2  + \left( {ch_{ + \mu }^{\left(  -  \right)}  - se^{ - i\omega } h_{ - \mu }^{\left(  -  \right)} } \right)^2 } \right)\tilde M_1^{ - 1}  \\ 
  + \left( {\left( {ch_{ - \mu }^{\left(  +  \right)}  + se^{i\omega } h_{ + \mu }^{\left(  +  \right)} } \right)^2  + \left( {ch_{ - \mu }^{\left(  -  \right)}  + se^{i\omega } h_{ + \mu }^{\left(  -  \right)} } \right)^2 } \right)\tilde M_2^{ - 1}  \\ 
 \end{array} \right],
\nonumber\\
&&
 M_{\mu \mu }^{\left(  -  \right)} {\rm{ }} =  - 2v^2 \left[ {\left( {ch_{ + \mu }^{\left(  +  \right)}  - se^{ - i\omega } h_{ - \mu }^{\left(  +  \right)} } \right)\left( {ch_{ + \mu }^{\left(  -  \right)}  - se^{ - i\omega } h_{ - \mu }^{\left(  -  \right)} } \right)\tilde M_1^{ - 1}  + \left( {ch_{ - \mu }^{\left(  +  \right)}  + se^{i\omega } h_{ + \mu }^{\left(  +  \right)} } \right)\left( {ch_{ - \mu }^{\left(  -  \right)}  + se^{i\omega } h_{ + \mu }^{\left(  -  \right)} } \right)\tilde M_2^{ - 1} } \right],
\nonumber\\
&&
 M_{\mu \tau }  =  - v^2 \left( { - \sigma } \right)\left[ \begin{array}{l}
 \left( {\left( {ch_{ + \mu }^{\left(  +  \right)}  - se^{ - i\omega } h_{ - \mu }^{\left(  +  \right)} } \right)^2  - \left( {ch_{ + \mu }^{\left(  -  \right)}  - se^{ - i\omega } h_{ - \mu }^{\left(  -  \right)} } \right)^2 } \right)\tilde M_1^{ - 1}  \\ 
  + \left( {\left( {ch_{ - \mu }^{\left(  +  \right)}  + se^{i\omega } h_{ + \mu }^{\left(  +  \right)} } \right)^2  - \left( {ch_{ - \mu }^{\left(  -  \right)}  + se^{i\omega } h_{ + \mu }^{\left(  -  \right)} } \right)^2 } \right)\tilde M_2^{ - 1}  \\ 
 \end{array} \right].
\label{Eq:NuMatrixEntries-2}
\end{eqnarray}
The suffices $\pm$ represent for $(N_+, N_-)$ that have $N_\pm\rightarrow N_\pm$ under the $\mu$-$\tau$ symmetry 
transformation.  The heavy neutrinos $(N_+, N_-)$ can be $(N_\mu, N_\tau)$, $(N_e, N_\mu)$ or any other combinations. 

\subsubsection{\label{appendix:1-2-1}$\mu$-$\tau$ Symmetry Breaking Case}
The approximate $\mu$-$\tau$ symmetry calls for 
\begin{eqnarray}
&& 
h_{+\mu }^{\left(  -  \right)}\approx 0, 
\quad
h_{- \mu }^{\left(  -  \right)}\approx 0, 
\label{Eq:FirstOrder3-1}
\end{eqnarray}
Using the approximation, we obtain from Eq.(\ref{Eq:NuMatrixEntries-2}) 
\begin{eqnarray}
&&
 M_{ee}  =  - v^2 \left[ {\left( {ch_{ + e}^{\left(  +  \right)}  - se^{ - i\omega } h_{ - e}^{\left(  +  \right)} } \right)^2 \tilde M_1^{ - 1}  + \left( {se^{i\omega } h_{ + e}^{\left(  +  \right)}  + ch_{ - e}^{\left(  +  \right)} } \right)^2 \tilde M_2^{ - 1} } \right],
\nonumber\\ 
&&
 M_{e\mu }^{\left(  +  \right)}  =  - v^2 \left[ {\left( {ch_{ + e}^{\left(  +  \right)}  - se^{ - i\omega } h_{ - e}^{\left(  +  \right)} } \right)\left( {ch_{ + \mu }^{\left(  +  \right)}  - se^{ - i\omega } h_{ - \mu }^{\left(  +  \right)} } \right)\tilde M_1^{ - 1}  + \left( {se^{i\omega } h_{ + e}^{\left(  +  \right)}  + ch_{ - e}^{\left(  +  \right)} } \right)\left( {se^{i\omega } h_{ + \mu }^{\left(  +  \right)}  + ch_{ - \mu }^{\left(  +  \right)} } \right)\tilde M_2^{ - 1} } \right],
\nonumber\\ 
&&
 M_{e\mu }^{\left(  -  \right)}  =  - v^2 \left[ {\left( {ch_{ + e}^{\left(  +  \right)}  - se^{ - i\omega } h_{ - e}^{\left(  +  \right)} } \right)\left( {ch_{ + \mu }^{\left(  -  \right)}  - se^{ - i\omega } h_{ - \mu }^{\left(  -  \right)} } \right)\tilde M_1^{ - 1}  + \left( {se^{i\omega } h_{ + e}^{\left(  +  \right)}  + ch_{ - e}^{\left(  +  \right)} } \right)\left( {se^{i\omega } h_{ + \mu }^{\left(  -  \right)}  + ch_{ - \mu }^{\left(  -  \right)} } \right)\tilde M_2^{ - 1} } \right],
\nonumber\\ 
&&
 M_{\mu \mu }^{\left(  +  \right)}  \approx  - v^2 \left[ {\left( {ch_{ + \mu }^{\left(  +  \right)}  - se^{ - i\omega } h_{ - \mu }^{\left(  +  \right)} } \right)^2 \tilde M_1^{ - 1}  + \left( {ch_{ - \mu }^{\left(  +  \right)}  + se^{i\omega } h_{ + \mu }^{\left(  +  \right)} } \right)^2 \tilde M_2^{ - 1} } \right],
\nonumber\\ 
&&
 M_{\mu \mu }^{\left(  -  \right)} {\rm{ }} =  - 2v^2 \left[ {\left( {ch_{ + \mu }^{\left(  +  \right)}  - se^{ - i\omega } h_{ - \mu }^{\left(  +  \right)} } \right)\left( {ch_{ + \mu }^{\left(  -  \right)}  - se^{ - i\omega } h_{ - \mu }^{\left(  -  \right)} } \right)\tilde M_1^{ - 1}  + \left( {ch_{ - \mu }^{\left(  +  \right)}  + se^{i\omega } h_{ + \mu }^{\left(  +  \right)} } \right)\left( {ch_{ - \mu }^{\left(  -  \right)}  + se^{i\omega } h_{ + \mu }^{\left(  -  \right)} } \right)\tilde M_2^{ - 1} } \right],
\nonumber\\ 
&&
 M_{\mu \tau }  \approx  - v^2 \left( { - \sigma } \right)\left[ {\left( {ch_{ + \mu }^{\left(  +  \right)}  - se^{ - i\omega } h_{ - \mu }^{\left(  +  \right)} } \right)^2 \tilde M_1^{ - 1}  + \left( {ch_{ - \mu }^{\left(  +  \right)}  + se^{i\omega } h_{ + \mu }^{\left(  +  \right)} } \right)^2 \tilde M_2^{ - 1} } \right],
\label{Eq:MassFlavorAppendix-2}
\end{eqnarray}
up to the first order in the parameters of Eq.(\ref{Eq:FirstOrder3-1}).

\subsubsection{\label{appendix:1-3-2}$\mu$-$\tau$ Symmetric Case}
We obtain from Eq.(\ref{Eq:NuMatrixEntries-2}) that 
\begin{eqnarray}
&&
 M_{ee} \left( { = a_0 } \right) =  - v^2 \left[ {\left( {ch_{ + e}^{\left(  +  \right)}  - se^{ - i\omega } h_{ - e}^{\left(  +  \right)} } \right)^2 \tilde M_1^{ - 1}  + \left( {se^{i\omega } h_{ + e}^{\left(  +  \right)}  + ch_{ - e}^{\left(  +  \right)} } \right)^2 \tilde M_2^{ - 1} } \right],
\nonumber\\ 
&&
 M_{e\mu } \left( { = b_0 } \right) = 
- v^2 \left[ \begin{array}{l}
  \left( {ch_{ + e}^{\left(  +  \right)}  - se^{ - i\omega } h_{ - e}^{\left(  +  \right)} } \right)\left( {ch_{ + \mu }^{\left(  +  \right)}  - se^{ - i\omega } h_{ - \mu }^{\left(  +  \right)} } \right)\tilde M_1^{ - 1}  \\ 
  + \left( {se^{i\omega } h_{ + e}^{\left(  +  \right)}  + ch_{ - e}^{\left(  +  \right)} } \right)\left( {se^{i\omega } h_{ + \mu }^{\left(  +  \right)}  + ch_{ - \mu }^{\left(  +  \right)} } \right)\tilde M_2^{ - 1}  \\ 
 \end{array} \right],
\nonumber\\ 
&&
 M_{e\tau } \left( { = c_0 } \right) =  - \sigma M_{e\mu },
\nonumber\\
&&
 M_{\mu \mu } \left( { = d_0  \equiv d_ +   + d_ -  } \right) =   - v^2 \left[ {\left( {ch_{ + \mu }^{\left(  +  \right)}  - se^{ - i\omega } h_{ - \mu }^{\left(  +  \right)} } \right)^2 \tilde M_1^{ - 1}  + \left( {ch_{ - \mu }^{\left(  +  \right)}  + se^{i\omega } h_{ + \mu }^{\left(  +  \right)} } \right)^2 \tilde M_2^{ - 1} } \right],
\nonumber\\ 
&&
 M_{\mu \tau } \left( { = e_0  \equiv  - \sigma \left( {d_ +   - d_ -  } \right) } \right) =  - \sigma M_{\mu \mu },
\nonumber\\ 
&&
 M_{\tau \tau } \left( { = f_0} \right) = M_{\mu \mu },
\label{Eq:NuMatrixEntriesSym-2}
\end{eqnarray}
leading $d_-=0$.  This texture gives the following mass terms:
\begin{eqnarray}
&&
\frac{1}{2}a_0 \nu _e \nu _e  + b_0 \frac{{\nu _e \nu _ +   + \nu _ +  \nu _e }}{{\sqrt 2 }} + d_ +  \nu _ +  \nu _ + .
\label{Eq:NuMassTermSym-2}
\end{eqnarray}
Since $m_3=0$ is realized because of the relation $e_0 = -\sigma d_0$ as in Eq.(\ref{Eq:PredictionsMassSym}), 
Eq.(\ref{Eq:NuMatrixEntriesSym-2}) is only consistent with the inverted mass hierarchy. 
Contrary to the previous case, the normal mass hierarchy is not 
realized in the minimal seesaw mechanism based on $N$ blind to the $\mu$-$\tau$ symmetry.

\section{\label{sec:Appendix2}Formula for Masses, Mixings and Phases.}
By adopting $U_{PMNS}$ of Eq.(\ref{Eq:Uu_general}) to diagonalize ${\rm\bf M}(\equiv M^\dagger_\nu M_\nu)$,  
where $U^\dagger_{PMNS}{\rm\bf M}U_{PMNS}$=diag.($m^2_1, m^2_2, m^2_3$) is satisfied, we can derive a set of
formula to express neutrino masses and mixing angles as well as phases in terms of the flavor neutrino masses 
 \cite{BabaYasue}.
The Hermitean matrix ${\rm\bf M}$ is parameterized by ${\rm\bf M}={\rm\bf M}^{(+)} +{\rm\bf M}^{(-)}$ with
\begin{eqnarray}
&&
{\rm\bf M}^{(+)}  = \left( \begin{array}{*{20}c}
   A & B_+ & - \sigma B_+  \\
   B^\ast_+ & D_+ & E_+   \\
   - \sigma B^\ast_+ & E_+ & D_+\\
\end{array} \right),
\quad
{\rm\bf M}^{(-)}  = \left( \begin{array}{*{20}c}
   0 & B_- & \sigma B_-  \\
   B^\ast_- & D_- & iE_-  \\
   \sigma B^\ast_- & -iE_- & - D_-\\
\end{array} \right),
\label{Eq:Mdagger-M}
\end{eqnarray}
where
\begin{eqnarray}
&&
A = \left| {M_{ee} } \right|^2  + 2\left( {\left| {M_{e\mu }^{( + )} } \right|^2  + \left| {M_{e\mu }
^{( - )} } \right|^2 } \right) ,
\nonumber \\ 
&&
B_ +   = M_{ee}^ \ast  M_{e\mu }^{( + )}  + M_{e\mu }^{( + ) \ast } \left( {M_{\mu \mu }^{( + )}  - 
\sigma M_{\mu \tau } } \right) + M_{e\mu }^{( - ) \ast } M_{\mu \mu }^{( - )}  ,
\nonumber \\ 
&&
B_ -   = M_{ee}^ \ast  M_{e\mu }^{( - )}  + M_{e\mu }^{( - ) \ast } \left( {M_{\mu \mu }^{( + )}  + 
\sigma M_{\mu \tau } } \right) + M_{e\mu }^{( + ) \ast } M_{\mu \mu }^{( - )},
\nonumber \\ 
&&
D_ +   = \left| {M_{e\mu }^{( + )} } \right|^2  + \left| {M_{e\mu }^{( - )} } \right|^2  + \left| {M_
{\mu \mu }^{( + )} } \right|^2  + \left| {M_{\mu \mu }^{( - )} } \right|^2  + \left| {M_{\mu \tau } }
 \right|^2 ,
\nonumber \\ 
&&
D_ -   = 2{\rm Re} \left( {M_{e\mu }^{( - ) \ast } M_{e\mu }^{( + )}  + M_{\mu \mu }^{( - ) \ast } M_
{\mu \mu }^{( + )} } \right),
\nonumber \\ 
&&
E_+ = {\rm Re}(E) = \sigma \left( {\left| {M_{e\mu }^{( - )} } \right|^2  - \left| {M_{e\mu }^{( + )} } \right|^2 }
 \right) + 2{\rm Re} \left( {M_{\mu \mu }^{( + ) \ast } M_{\mu \tau } } \right),
\nonumber\\
&&
E_- = {\rm Im}(E) = 2{\rm Im} \left( {M_{\mu \mu }^{( - ) \ast } M_{\mu \tau }  - \sigma M_{e\mu }^{( - ) \ast } M_
{e\mu }^{( + )} } \right),
\label{Eq:A-F}
\end{eqnarray}
for $E=E_++iE_-$. Similarly, we define 
$B$=$B_++B_-$, $C$=$-\sigma (B_+-B_-)$, $D$=$D_++D_-$, and $F$=$D_+-D_-$ to describe matrix elements of ${\rm\bf M}$. 
We, then, obtain that
\begin{eqnarray}
&&
\tan 2\theta _{12}e^{i\rho}  = \frac{{2X}}{{\Lambda _2  - \Lambda_1 }},
\quad
\tan 2\theta _{13}e^{-i\delta}  = \frac{2Y}{{\Lambda _3  - A}},
\label{Eq:Mdagger-ExactMixingAngles12-13} 
\nonumber\\
&&
{\rm Re} \left( {e^{ - 2i\gamma } E} \right)\cos 2\theta _{23}  + D_-\sin 2\theta _{23}
  + i{\rm Im} \left( {e^{ - 2i\gamma } E} \right) =  - s_{13} e^{- i\delta} X^\ast,
\label{Eq:Mdagger-ExactMixingAngles23}
\end{eqnarray}
for three mixing angles and three phases, and
\begin{eqnarray}
&&
m_1^2  = c_{12}^2 \Lambda _1  + s_{12}^2 \Lambda _2  - 2c_{12} s_{12} \left| X\right|,
\quad
m_2^2  = s_{12}^2 \Lambda _1  + c_{12}^2 \Lambda _2  + 2c_{12} s_{12} \left| X\right|,
\quad
m_3^2  = \frac{{c_{13}^2 \Lambda _3  - s_{13}^2 A}}{{c_{13}^2  - s_{13}^2 }}, 
\label{Eq:Mdagger-ExactMasses}
\end{eqnarray}
for three masses, where
\begin{eqnarray}
&&
X = \frac{{c_{23} e^{i\gamma} B - s_{23} e^{-i\gamma} C}}{c_{13} } 
= \frac{e^{i\rho}\left| c_{23} e^{i\gamma} B - s_{23} e^ {- i\gamma} C\right|}{c_{13} },
\nonumber\\
&&
Y = s_{23} e^{i\gamma } B + c_{23} e^{ - i\gamma } C
= e^{-i\delta}\left| s_{23} e^{i\gamma } B + c_{23} e^{ - i\gamma } C\right|,
\nonumber\\
&&
\Lambda _1  = \frac{{c_{13}^2 A - s_{13}^2 \Lambda _3 }}{{c_{13}^2  - s_{13}^2 }},
\quad
\Lambda _2  = c_{23}^2 D + s_{23}^2 F - 2s_{23} c_{23} {\rm Re} \left( {e^{ - 2i\gamma } E} \right),
\quad
\nonumber\\
&&
\Lambda _3  = s_{23}^2 D + c_{23}^2 F + 2s_{23} c_{23} {\rm Re} \left( {e^{ - 2i\gamma } E} \right).
\label{Eq:Mdagger-MassParameters}
\end{eqnarray}
In the $\mu$-$\tau$ symmetric case, where $B_-=D_-=E_-=0$, we obtain that $\rho=\arg(B)$, 
$\gamma = 0$, and $\cos 2\theta_{23} = \sin\theta_{13}=0$.
The Dirac CP violation involves the angle $\rho+\delta$. 

There are useful relations:
\begin{eqnarray}
&&
\left| X\right|= \frac{\Delta m^2_\odot\sin 2\theta_{12}}{2},
\quad
 A \approx 
\frac{{\Sigma m_ \odot ^2  - \cos 2\theta _{12} \Delta m_ \odot ^2  + s_{13}^2 \left( {2\Delta m_{atm}^2  - \left( {1 - \cos 2\theta _{12} } \right)\Delta m_ \odot ^2 } \right)}}{2},
\nonumber\\
&&
 D_ +   \approx 
\frac{1}{2}\left( {\Delta m_{atm}^2  + \Sigma m_ \odot ^2  - \frac{{\left( {1 - \cos 2\theta _{12} } \right)\Delta m_ \odot ^2  + s_{13}^2 \left( {2\Delta m_{atm}^2  - \left( {1 - \cos 2\theta _{12} } \right)\Delta m_ \odot ^2 } \right)}}{2}} \right),
\nonumber\\
&&
 \sigma {\rm Re} \left( {e^{ - 2i\gamma } E} \right) - 2D_ -  \Delta  \approx 
\frac{1}{2}\left( {\Delta m_{atm}^2  - \frac{{\left( {1 + \cos 2\theta _{12} } \right)\Delta m_ \odot ^2  + s_{13}^2 \left( {2\Delta m_{atm}^2  - \left( {1 - \cos 2\theta _{12} } \right)\Delta m_ \odot ^2 } \right)}}{2}} \right),
\nonumber\\
&&
 \Lambda _1  \approx \frac{{\Sigma m_ \odot ^2  - \cos 2\theta _{12} \Delta m_ \odot ^2 }}{2},
\quad
 \Lambda _2 = \frac{{\cos 2\theta _{12} \Delta m_ \odot ^2  + \Sigma m_ \odot ^2 }}{2},
\nonumber\\
&&
 \Lambda _3  \approx 
\frac{{2\Delta m_{atm}^2  + \Sigma m_ \odot ^2  - \Delta m_ \odot ^2  - s_{13}^2 \left( {2\Delta m_{atm}^2  - \left( {1 - \cos 2\theta _{12} } \right)\Delta m_ \odot ^2 } \right)}}{2},
\label{Eq:Mdagger-M-MassRelation}
\end{eqnarray}
up to ${\mathcal{O}}(\sin^2\theta_{13})$, where $\sum m_ \odot ^2  = m_1^2  + m_2^2$.  
The real part of Eq.(\ref{Eq:Mdagger-ExactMixingAngles23})
\begin{eqnarray}
&&
{\rm Re} \left( {e^{ - 2i\gamma } E} \right)\cos 2\theta _{23}  + D_ -  \sin 2\theta _{23} 
=  -s_{13} \cos \left(\rho  + \delta \right) \left|X\right|\left( \equiv -z\right),
\label{Eq:Exact-Angle23}
\end{eqnarray}
determines $\cos 2\theta_{23}$, which is given by
\begin{eqnarray}
&&
\cos 2\theta _{23}  =  -  \frac{{\kappa\sigma D_ -\sqrt {{\rm Re} ^2 \left( {e^{ - 2i\gamma } E} \right) + D_ - ^2  - z^2 }   + z{\rm Re} \left( {e^{ - 2i\gamma } E} \right)}}{{{\rm Re} ^2 \left( {e^{ - 2i\gamma } E} \right) + D_ - ^2 }}
=  \cos \left( \sigma\frac{\pi}{2}+\theta + \phi \right),
\nonumber\\
&&
\cos \theta  = \sqrt {\frac{{\rm Re} ^2 \left( {e^{ - 2i\gamma } E} \right) + D_ - ^2  - z^2 }{{\rm Re} ^2 \left( {e^{ - 2i\gamma } E} \right) + D_ - ^2  }},
\quad
\sin \theta  = \frac{\sigma z}{\sqrt {{\rm Re} ^2 \left( {e^{ - 2i\gamma } E} \right) + D_ - ^2 } },
\nonumber\\
&&
\cos \phi  = \frac{{\rm Re} \left( {e^{ - 2i\gamma } E} \right)}{\sqrt {{\rm Re} ^2 \left( {e^{ - 2i\gamma } E} \right) + D_ - ^2 } },
\quad
\sin \phi  = \frac{\kappa D_-}{\sqrt {{\rm Re} ^2 \left( {e^{ - 2i\gamma } E} \right) + D_ - ^2 } },
\label{Eq:Exact-cos23-solution}
\end{eqnarray}
where $\kappa$ is the sign of ${\rm Re}( e^{ - 2i\gamma }E)$, from which we obtain that 
$\theta _{23} = \sigma\pi/4 + (\theta + \phi)/2$.
On the other hand, the imaginary part of Eq.(\ref{Eq:Mdagger-ExactMixingAngles23}) 
\begin{eqnarray}
&&
\cos 2\gamma {\rm Im} \left( E \right) - \sin 2\gamma {\rm Re}\left( E \right) = s_{13} \sin \left( {\rho  + \delta } \right)\left|X\right|\left( \equiv z^\prime\right),
\label{Eq:Exact-gamma}
\end{eqnarray}
determines $\gamma$, which is given by
\begin{eqnarray}
&&
{\sin 2\gamma  = \frac{{ \kappa^\prime {\rm Im}\left( E \right)\sqrt {\left| E \right|^2  - z^{\prime 2} } - z^\prime {\rm Re} \left( E \right)}}{{\left| E \right|^2 }} 
=  \sin \left( \phi^\prime-\theta^\prime \right)},
\nonumber\\
&&
\cos \theta^\prime  = \frac{\sqrt {\left| E \right|^2  - z^{\prime 2} }}{{\left| E \right|}},
\quad
\sin \theta^\prime  = \frac{z^\prime}{{\left| E \right|}},
\nonumber\\
&&
\cos \phi^\prime  = \frac{{\rm Re} \left( E \right)}{{\left| E \right|}},
\quad
\sin \phi^\prime  = \frac{\kappa^\prime{\left| {{\rm Im} \left( E \right)} \right|}}{{\left| E \right|}},
\label{Eq:Exact-gamma-solution}
\end{eqnarray}
where $\kappa^\prime$ is the sign of ${\rm Re}( E)$, from which we obtain that 
$\gamma = (\phi^\prime-\theta^\prime)/2 $.

The Majorana phases are calculated by the following formula derived by 
$U_{PMNS}^T M_\nu U_{PMNS}$ = diag.$(m_1,m_2,m_3)$:
\begin{eqnarray}
&&
 m_1 e^{ - 2i\left(\phi^\prime_1-\rho\right) } 
 \left( = m_1 e^{ - 2i\phi_1 }\right)
 = \frac{{\lambda _1  + \lambda _2 }}{2} - \frac{x}{{\sin 2\theta _{12} }}, 
 \quad 
 m_2 e^{ - 2i\phi_2 }  = \frac{{\lambda _1  + \lambda _2 }}{2} + \frac{x}{{\sin 2\theta _{12} }},
\nonumber\\
&&
 m_3 e^{ - 2i\phi_3 }  =  \frac{{c_{13}^2 \lambda _3  - s_{13}^2e^{-2i\delta } a}}{{c_{13}^2  - s_{13}^2 }},
\label{Eq:M}
\end{eqnarray}
where
\begin{eqnarray}
&&
 \lambda _1  = e^{2i\rho } \frac{{c_{13}^2 a - s_{13}^2 e^{  2i\delta} \lambda _3 }}{{c_{13}^2  - s_{13}^2 }},
\quad
 \lambda _2  = c_{23}^2 e^{2i\gamma } d + s_{23}^2 e^{ - 2i\gamma } f - 2s_{23} c_{23} e,
\nonumber\\
&&
 \lambda _3  = s_{23}^2 e^{2i\gamma } d + c_{23}^2 e^{ - 2i\gamma } f + 2s_{23} c_{23} e,
\quad
 x = \frac{{e^{i\rho } \left( {c_{23} e^{i\gamma } b - s_{23} e^{ - i\gamma } c} \right)}}{{c_{13} }}.
\label{Eq:M-lambda}
\end{eqnarray}
The CP violating Majorana phase denoted by $\phi$ is represented by $(\phi_2-\phi_3)/2$ 
 for $m_1=0$, leading to $K$ = diag.$(1,e^{i\phi}, e^{-i\phi})$, and
by $ (\phi_1-\phi_2)/2$  for $m_3=0$, leading to $K$ = diag.$(e^{i\phi}, e^{-i\phi}, 1)$.
To see the phase of $M_{ee}$, which affects the detection of the absolute neutrino mass $m_{\beta\beta}$ 
in double beta decay experiments, 
we have to refer to $U_{PMNS}$ of Eq.(\ref{Eq:Uu}) denoted by $U^{PDG}_{PMNS}$, which is associated with 
$M^{PDG}_\nu$ defined by
\begin{eqnarray}
&&
U^{PDG~T}_{PMNS} M^{PDG}_\nu U^{PDG}_{PMNS} = U_{PMNS}^T M_\nu U_{PMNS},
\label{Eq:U-M-PDG}
\end{eqnarray}
where $M^{PDG}_\nu$ is used in theoretical calculations compared with results of neutrino experiments.  
From Eq.(\ref{Eq:U-M-PDG}), by adjusting phases of the flavor neutrinos 
\begin{eqnarray}
&&
\nu '_L  = \left( {\begin{array}{*{20}c}
   {e^{ - i\rho } } & 0 & 0  \\
   0 & {e^{ - i\gamma } } & 0  \\
   0 & 0 & {e^{i\gamma } }  \\
\end{array}} \right)\nu _L ,
\label{Eq:nu-redifined}
\end{eqnarray}
for $\nu _L = (\nu_e, \nu_\mu. \mu_\tau)^T$ used in Eq.(\ref{Eq:SuperPotential}), we find that
\begin{eqnarray}
&&
M^{PDG}_\nu = \left( {\begin{array}{*{20}c}
	e^{2i\rho }M_{ee} & e^{i\left( {\rho  + \gamma } \right)}M_{e\mu} & e^{i\left( {\rho - \gamma } \right)}M_{e\tau}  \\
	e^{i\left( {\rho  + \gamma } \right)}M_{e\mu} & e^{2i\gamma }M_{\mu\mu} & M_{\mu\tau}  \\
	e^{i\left( {\rho - \gamma } \right)}M_{e\tau} & M_{\mu\tau} & e^{-2i\gamma }M_{\tau\tau}  \\
\end{array}} \right),
\nonumber \\
&&
K^{PDG} = {\rm diag}(e^{i\phi_1}, e^{i\phi_2}, e^{i\phi_3}), \quad \delta_{CP}=\delta+\rho, \quad \phi_1 = \phi^\prime_1-\rho,
\label{Eq:M-nu-PDG}
\end{eqnarray}
where $K^{PDG}$ is obtained from $K = {\rm diag}(e^{i\phi^\prime_1}, e^{i\phi_2}, e^{i\phi_3})$,
as defined in Eq.(\ref{Eq:Uu_general}).  
Therefore, it should be noted that $m_{\beta\beta}$ is equal to
\begin{eqnarray}
&&
e^{2i\rho }M_{ee},
\label{Eq:M-nu-ee-PDG}
\end{eqnarray}
but not to $M_{ee}$.


\newpage
\noindent
\noindent
\noindent
\end{document}